\documentclass[11pt,a4paper]{article}
\pdfoutput=1

\usepackage{jheppub}

\usepackage{amsmath}
\usepackage{bbm}
\usepackage{float}
\usepackage{graphicx}	
\usepackage{hyperref}
\usepackage{mathtools}
\usepackage{cancel}

\usepackage[normalem]{ulem}

\usepackage{multirow}
\usepackage{rotating}
\usepackage{subcaption}

\def\lsim{\raise0.3ex\hbox{$\;<$\kern-0.75em\raise-1.1ex
\hbox{$\sim\;$}}}
\def\gsim{\raise0.3ex\hbox{$\;>$\kern-0.75em\raise-1.1ex
\hbox{$\sim\;$}}}

\def\thetitle{An Effective Two-Flavor Approximation for Neutrino Survival Probabilities in Matter 
 {\small \vskip -4.6cm \hglue 11cm \rm YACHAY-PUB-17-02-PN}
 \vspace{2.5cm}
}
\title{\thetitle}
\hypersetup{pdftitle={\thetitle}}
\author{Hisakazu Minakata}
\affiliation{
Department of Physics, Yachay Tech, San Miguel de Urcuqu\'i, 100119 Ecuador \\
}

\abstract{ 
It is known in vacuum that the three-flavor neutrino survival probability can be approximated by the effective two-flavor form to first orders in $\epsilon \equiv \Delta m^2_{21} / \Delta m^2_{31}$, with introduction of the effective $\Delta m^2_{\alpha \alpha}$ ($\alpha = e, \mu, \tau$), in regions of neutrino energy $E$ and baseline $L$ such that $\Delta m^2_{31} L / 2E \sim \pi$. Here, we investigate the question of whether the similar effective two-flavor approximation can be formulated for the survival probability in matter. Using a perturbative framework with the expansion parameters $\epsilon$ and $s_{13} \propto \sqrt{\epsilon}$, we give an affirmative answer to this question and the resultant two-flavor form of the probability is valid to order $\epsilon$. However, we observe a contrived feature of the effective $\Delta m^2_{\alpha \alpha} (a)$ in matter. It ceases to be a combination of the fundamental parameters and has energy dependence, which may be legitimate because it comes from the matter potential. But, it turned out that $\Delta m^2_{\mu \mu} (a)$ becomes $L$-dependent, though $\Delta m^2_{ee} (a)$ is not, which casts doubt on adequacy of the concept of effective $\Delta m^2$ in matter. We also find that the appearance probability in vacuum admits, to order $\epsilon$, the similar effective two-flavor form with a slightly different effective $\Delta m^2_{\beta \alpha}$ from the disappearance channel. 
A general result is derived to describe suppression of the matter effect in the oscillation probability. 

}

\emailAdd{hminakata@yachaytech.edu.ec}

\begin{document} 

\maketitle

\section{Introduction}

After great success of the three-flavor mixing scheme of neutrinos describing almost all data available to date, the neutrino experiments entered into the era of precision measurement and paradigm test. Here, it may be interesting to pay attention to the mutually different roles played by the appearance and the disappearance channels. The appearance channel $\nu _\mu \rightarrow \nu _e$ (or its T-conjugate) can play an important role to signal new effects, such as giving the first indication of nonzero $\theta_{13}$ \cite{Abe:2011sj}, which would also offer the best chance for discovering lepton CP violation in the future \cite{Abe:2015zbg,Acciarri:2015uup}. 
On the other hand, precision measurements of the mixing parameters to date are carried out mostly by using the disappearance channels $\nu _\alpha \rightarrow \nu _\alpha$ ($\alpha = e, \mu$ including antineutrino channels). It includes tens of experiments using the atmospheric, solar, reactor and the accelerator neutrinos as in e.g., \cite{SK-neutrino-2016,Aharmim:2011vm,Abe:2016nxk,Gando:2013nba,An:2016ses,RENO:2015ksa,Abe:2014bwa,Adamson:2014vgd,Abe:2015awa,NOvA-neutrino-2016}, which are on one hand complementary to each other, but on the other hand are competing toward the best accuracy. 
To quote another example of their complementary roles, precision measurement of the survival probabilities in the channels $\nu _{e} \rightarrow \nu_{e}$ and $\nu _\mu \rightarrow \nu _\mu$ is essential for accurate determination of $\theta_{13}$ and $\theta_{23}$, respectively, while the appearance channel helps as a degeneracy solver \cite{Minakata:2002jv,Hiraide:2006vh}. Therefore, precise knowledge of the disappearance channel oscillation probability in matter could be of some help for a better understanding of the data with diverse experimental settings. 
Hereafter, we refer $\nu _\alpha$ disappearance channel oscillation probability $P(\nu _\alpha \rightarrow \nu _\alpha)$ as the $\nu _\alpha$ survival probability. 

In this context, it is noteworthy that the authors of ref.~\cite{Nunokawa:2005nx} presented effective two-flavor description of the three-flavor neutrino survival probability in vacuum. They introduced an effective $\Delta m^2$ to describe superposition of the atmospheric-scale oscillations with the two different frequencies associated with $\Delta m^2_{32}$ and $\Delta m^2_{31}$. Interestingly, the effective $\Delta m^2$ is channel dependent: $\Delta m^2_{ee} = c^2_{12} \Delta m^2_{31} + s^2_{12} \Delta m^2_{32}$ and $\Delta m^2_{\mu \mu} = s^2_{12} \Delta m^2_{31} + c^2_{12} \Delta m^2_{32}$, respectively, to zeroth order in $\sin \theta_{13}$. It suggests a possibility that the effective $\Delta m^2$ measured in the reactor $\bar{\nu}_{e}$ \cite{An:2016ses,RENO:2015ksa} (see \cite{An:2013zwz} for the first measurement) and the accelerator $\nu_{\mu}$ \cite{Adamson:2014vgd,Abe:2015awa,NOvA-neutrino-2016} disappearance experiments can have a tiny difference of the order of $\Delta m^2_{21}$. If observed, the difference between $\Delta m^2_{ee}$ and $\Delta m^2_{\mu \mu}$ could have an important implication because the sign of $\Delta m^2_{ee} - \Delta m^2_{\mu \mu}$ will tell us about which neutrino mass ordering is chosen by nature \cite{Nunokawa:2005nx,Minakata:2006gq}. 

In this paper, we investigate the question of whether the similar effective two-flavor description of the three-flavor neutrino survival probability is viable for neutrinos propagating in matter. We emphasize that it is a highly nontrivial question because the structure of neutrino oscillations is drastically altered in the presence of Wolfenstein's matter potential $a$ in the Hamiltonian \cite{Wolfenstein:1977ue}. It also brings a different (not in the form of $1/E$) energy dependence into the Hamiltonian. Using perturbative expression of the survival probability $P(\nu_\alpha \rightarrow \nu_\alpha)$ in matter, and by introducing the similar ansatz for the effective two-flavor form of the probability as in vacuum, we will give an affirmative answer to the question to first order in the small expansion parameter $\epsilon \equiv \Delta m^2_{21} / \Delta m^2_{31}$. The ansatz includes the effective two-flavor $\Delta m^2_{\alpha \alpha} (a)$ ($\alpha=e, \mu, \tau$) in matter as a natural generalization of $\Delta m^2_{\alpha \alpha}$ in vacuum. 

But, then, it turned out that $\Delta m^2_{\alpha \alpha} (a)$ becomes a dynamical quantity, which depends on neutrino energy $E$. It may be inevitable and legitimate because the energy dependence comes in through the matter potential $a \propto E$. However, a contrived feature appears in $\Delta m^2_{\mu \mu} (a)$ that it depends on $L$, the baseline distance. This feature does not show up in $\Delta m^2_{ee} (a)$. Thus, while the effective two-flavor description of the three-flavor neutrino survival probability in matter seems to be possible, the resultant effective $\Delta m^2_{\alpha \alpha} (a)$ does not appear to possess any fundamental significance as a physical parameter. We will argue that this feature is not due to the artifact of the perturbative treatment. 

Let us start by refreshing our understanding of the effective two-flavor description of the three-flavor neutrino survival probability in vacuum. 

\section{Validity of the effective two-flavor approximation in vacuum}
\label{sec:validity}

Suppose that one can measure neutrino energy with an extreme precision, $\frac{\Delta E}{E} \ll \frac{\Delta m^2_{21}}{\Delta m^2_{31}}$.\footnote{
This condition is derived by requiring uncertainty of the kinematical factor $\frac{\Delta m^2_{31}L}{4E}$ of $\Delta m^2_{31}$ wave due to energy resolution $\Delta E$ is much smaller than the difference between the $\Delta m^2_{31}$ and $\Delta m^2_{32}$ waves, $\frac{\Delta m^2_{21}L}{4E}$.
}
Let us then ask a question: 
Can one observe two dips in the energy spectrum of $\nu_{\mu}$ in muon neutrino disappearance measurement due to two waves modulated with two different frequencies associated with $\Delta m^2_{32}$ and $\Delta m^2_{31}$? In vacuum and at around the first oscillation maximum (i.e., highest-energy maximum) of the atmospheric scale oscillation, $\frac{ \Delta m^2_{31} L }{ 2E} \simeq \pi$, we can give a definitive answer to the question; one never. It will be demonstrated below. If the same feature holds in matter, it provides us the raison d'~\^etre for the approximate effective two-flavor form for the survival probability in matter in the three-flavor mixing scheme. 

In the rest of this section, we start from the ``proof'' showing that in vacuum the $\Delta m^2_{32}$ and $\Delta m^2_{31}$ waves always form a single collective wave and has no chance to develop two minima in the energy spectrum of survival probability $P(\nu_\alpha \rightarrow \nu_\alpha)$, where $\alpha$ is one of $e$, $\mu$, or $\tau$. 
Then, we formulate an ansatz for the effective two-flavor approximation of the three-flavor probabilities in vacuum, which in fact gives a premise for the similar treatment in matter. 

\subsection{Two waves form a single collective wave in vacuum}
\label{sec:two-wave}

We discuss the $\nu_{\alpha}$ survival probability $P(\nu_\alpha \rightarrow \nu_\alpha)$ ($\alpha=e, \mu, \tau$) in vacuum to understand the reasons why we expect that the effective two-flavor approximation is valid. Using unitarity, it can be written without any approximation as \cite{Minakata:2007tn}
\begin{eqnarray}
P(\nu_\alpha \rightarrow \nu_\alpha) &=& 1- 
4\vert U_{\alpha 3}\vert^2 \vert U_{\alpha 1}\vert^2 \sin^2 \Delta_{31} -
4 \vert U_{\alpha 3}\vert^2 \vert U_{\alpha 2}\vert^2 \sin^2 \Delta_{32}  -
4\vert U_{\alpha 2}\vert^2 \vert U_{\alpha e1}\vert^2 \sin^2 \Delta_{21}, 
\nonumber \\
&=& 
1 - 4\vert U_{\alpha 2}\vert^2 \vert U_{\alpha e1}\vert^2 \sin^2 \Delta_{21} 
\nonumber \\
&-&
2 \vert U_{\alpha 3}\vert^2 
\left( \vert U_{\alpha 1}\vert^2 + \vert U_{\alpha 2}\vert^2 \right) 
\left[
1 - \sqrt{1-\sin^2 2 \chi \sin^2 \Delta_{21}} 
~\cos (2 \Delta_{\alpha \alpha} \pm \phi) 
\right]
\label{P-alpha-alpha-vac}
\end{eqnarray}
where the sign $\pm$ in the cosine function at the end correspond to the mass ordering, $+$ for the normal and $-$ for inverted orderings. $U_{\alpha j}$ $(j=1,2,3)$ denotes the MNS matrix elements \cite{Maki:1962mu}. The kinematical factor $\Delta_{j i}$ used in eq.~(\ref{P-alpha-alpha-vac}) is defined as 
\begin{eqnarray}
\Delta_{j i} \equiv \frac{\Delta m^2_{j i} L }{4E}, 
\hspace{10mm}
(i, j = 1, 2, 3), 
\label{Delta-ji-def}
\end{eqnarray}
where $E$ is neutrino energy and $L$ the baseline distance. $\Delta m^2_{ji}$ denote neutrino mass squared differences, $\Delta m^2_{ji} \equiv m^2_{j} - m^2_{i}$ $(i, j = 1,2,3)$.

The angle $\chi$ in the square root in (\ref{P-alpha-alpha-vac}) are defined  as
\begin{eqnarray}
\cos \chi = \frac{ \vert U_{\alpha 1} \vert }{ \sqrt{ \vert U_{\alpha 1}\vert^2 + \vert U_{\alpha 2}\vert^2  } },
\hspace{10mm}
\sin \chi = \frac{ \vert U_{\alpha 2} \vert }{ \sqrt{ \vert U_{\alpha 1}\vert^2 + \vert U_{\alpha 2}\vert^2  } }.
\label{chi-def}
\end{eqnarray}
Now, $\Delta_{\alpha \alpha}$ in the argument of the cosine function in (\ref{P-alpha-alpha-vac}) is defined as follows: 
\begin{eqnarray}
\Delta_{\alpha \alpha} \equiv \frac{\Delta m^2_{\alpha \alpha} L }{4E}, 
\hspace{10mm}
\Delta m^2_{\alpha \alpha} \equiv \cos^2 \chi \vert \Delta m^2_{31} \vert + \sin^2 \chi \vert \Delta m^2_{32} \vert.
\label{dm2-mumu}
\end{eqnarray}
Finally, the phase $\phi$ is defined as
\begin{eqnarray}
\cos \phi & = & 
\frac{\cos^2 \chi \cos \left( 2 \sin^2 \chi \Delta_{21} \right) + \sin^2 \chi \cos \left( 2 \cos^2 \chi \Delta_{21} \right) }
{ \sqrt{1-\sin^2 2 \chi \sin^2 \Delta_{21}}},
\nonumber \\
\sin \phi  & = & 
\frac{ \cos^2 \chi \sin \left( 2 \sin^2 \chi \Delta_{21} \right)
- \sin^2 \chi \sin \left( 2 \cos^2 \chi \Delta_{21} \right) }
{\sqrt{1-\sin^2 2 \chi \sin^2 \Delta_{21}}}. 
\label{phi-def}
\end{eqnarray}
Notice that $\phi$ depends only on the 1-2 sector variables, or the ones relevant for the solar-scale oscillations.

Thanks to the hierarchy of the two $\Delta m^2$, 
\begin{eqnarray}
\epsilon \equiv \frac{\Delta m^2_{21}}{\Delta m^2_{31}} \approx 0.03 \ll 1, 
\label{epsilon-def}
\end{eqnarray}
one can obtain a perturbative expression of $\sin \phi$, 
\begin{eqnarray}
\sin \phi = \frac{ \epsilon^3 }{3} \sin^2 2\chi \cos 2\chi (\Delta_{31})^3 + \mathcal{O} (\epsilon^5), 
\label{phi-approx}
\end{eqnarray}
which shows that $\sin \phi$ is extremely small, $\sin \phi \lsim 10^{-5}$, at around the first oscillation maximum of atmospheric scale oscillations, $\Delta_{31} \sim 1$. (The similar argument applies also to the second oscillation maximum.) Notice that at $\Delta_{21} = \epsilon \Delta_{31} \sim 1$, the perturbative expansion breaks down. 

Thus, the superposed wave in the last line in (\ref{P-alpha-alpha-vac}) can be well approximated by a single harmonic and there is no way that $\Delta_{31}$ and $\Delta_{32}$ waves develop two minima inside the region of interest, $0 < \Delta_{31} \sim \Delta_{32} < \pi$. Notice that the modulation due to the solar $\Delta m^2_{21}$ term in (\ref{P-alpha-alpha-vac}) does not alter this conclusion because of its much longer wavelength by a factor of $\sim 30$.

One may argue that were the baseline $\Delta_{21} \sim 1$ is used instead, then one can distinguish between oscillations due to $\Delta m^2_{31}$ and $\Delta m^2_{32}$ waves, thereby could see the double dips. Despite that the former statement is in a sense true, the latter is not. In other word, what happens is different in nature. The feature that the superposed two waves behave as a single harmonics prevails. The difference between $\Delta m^2_{31}$ and $\Delta m^2_{32}$, $\vert \Delta m^2_{31} \vert > \vert \Delta m^2_{32} \vert$ (normal mass ordering) or $\vert \Delta m^2_{31} \vert < \vert \Delta m^2_{32} \vert$ (inverted mass ordering), entails advancement or retardation of the phase of the single wave formed by superposition \cite{Minakata:2007tn}. Therefore, it appears that the property of no double dip generically applies even in the case $\Delta_{21} \sim \Delta_{31}$. However, we did not try to make the statement of no double dip in $P(\nu_\alpha \rightarrow \nu_\alpha)$ in vacuum at all energies and the whole parameter regions a rigorous theorem.

\subsection{Effective two-flavor approximation in vacuum}
\label{sec:2flavor-vac}

In this section, we try to provide the readers a simpler way of understanding the results obtained in ref.~\cite{Nunokawa:2005nx}. We postulate the following ansatz for an effective two-flavor form of the three-flavor $\nu _\alpha$ survival probability $P(\nu _\alpha \rightarrow \nu _\alpha)$ ($\alpha= e, \mu, \tau$) in vacuum which is valid up to order $\epsilon$, 
\begin{eqnarray} 
P(\nu _\alpha \rightarrow \nu _\alpha) &=&
C_{\alpha \alpha} - A_{\alpha \alpha} \sin ^2 \left(\frac{\Delta m^2_{\alpha \alpha} L}{4E} \right). 
\label{P-alpha-alpha-vacuum-ansatz}
\end{eqnarray}
In principle, it is also possible to seek the effective two-flavor form which is valid to higher order in $\epsilon$ by adopting more complicated ansatz. But, we do not try to pursue this line in this paper to keep the simplicity of the resultant expressions. We remark that, throughout this paper, we limit ourselves into the region $\Delta_{31} \sim \Delta_{32} \sim \Delta_{\alpha \alpha} \sim 1$ for the effective two-flavor formulas to work both in vacuum and in matter. Therefore, $\Delta_{21}$ is of the order of $\epsilon$. 

For clarity we discuss here a concrete example, $P(\nu _\mu \rightarrow \nu _\mu)$ in vacuum. In this paper we use the PDG parametrization of the MNS matrix. We keep the terms of order $\Delta_{21}^2 \sim \epsilon^2$, the only exercise we engage in this paper to examine the order $\epsilon^2$ terms. It is to give a feeling to the readers on how the two-flavor ansatz could (or could not) be extended to order $\epsilon^2$. 
The $\nu_{\mu}$ survival probability $P(\nu _\mu \rightarrow \nu _\mu)$ in vacuum can be written to second order in $\epsilon$ as 
\begin{eqnarray}
&& P(\nu _\mu \rightarrow \nu _\mu) 
\nonumber \\
&=& 
1 - 4 \epsilon^2 \left(s^2_{12} c^2_{23} + c^2_{12} s^2_{23} s^2_{13} + 2 J_r \cos \delta \right) \left(c^2_{12} c^2_{23} + s^2_{12} s^2_{23} s^2_{13} - 2 J_r \cos \delta \right) \Delta_{31}^2
\nonumber \\
&-& 4 s^2_{23} c^2_{13} \left( c^2_{23} + s^2_{23} s^2_{13} \right) \sin ^2 \Delta_{31} 
\nonumber \\
&+& 4 \epsilon s^2_{23} c^2_{13} 
\left(c^2_{12} c^2_{23} + s^2_{12} s^2_{23} s^2_{13} - 2 J_r \cos \delta \right)
\Delta_{31} \sin 2\Delta_{31} 
\nonumber \\
&-& 
4 \epsilon^2 s^2_{23} c^2_{13} 
\left(c^2_{12} c^2_{23} + s^2_{12} s^2_{23} s^2_{13} - 2 J_r \cos \delta \right) \Delta_{31}^2 \cos 2\Delta_{31},
\label{Pmumu-vac2}
\end{eqnarray}
where $J_r \equiv c_{12} s_{12} c_{23} s_{23} s_{13}$. 

We examine whether a simple ansatz for $\Delta m^2_{\alpha \alpha}$ in (\ref{P-alpha-alpha-vacuum-ansatz}), 
\begin{eqnarray}
\Delta m^2_{\alpha \alpha} = \Delta m^2_{31} - s_{\alpha} \Delta m^2_{21}, 
\label{Dm2-eff-ansatz-vac}
\end{eqnarray}
can be matched to (\ref{Pmumu-vac2}) to order $\epsilon$.
The $\nu_{\alpha}$ survival probability $P(\nu _\alpha \rightarrow \nu _\alpha) $ in (\ref{P-alpha-alpha-vacuum-ansatz}) can be expanded to a power series of $\Delta_{21}$ as 
\begin{eqnarray}
&& P(\nu _\alpha \rightarrow \nu _\alpha) =
C_{\alpha \alpha} - A_{\alpha \alpha} \sin^2 \Delta_{31} 
+ A_{\alpha \alpha} (s_{\alpha} \Delta_{21}) \sin 2\Delta_{31} 
- A_{\alpha \alpha} (s_{\alpha} \Delta_{21})^2 \cos 2 \Delta_{31}.  
\nonumber \\
\label{Pmumu-2flavor-2nd}
\end{eqnarray}
The equations (\ref{Pmumu-vac2}) and (\ref{Pmumu-2flavor-2nd}) matches (for $\alpha=\mu$) to order $\Delta_{21}$ if  
\begin{eqnarray}
C_{\mu \mu} &=& 1 - 4 \epsilon^2 \left(s^2_{12} c^2_{23} + c^2_{12} s^2_{23} s^2_{13} + 2 J_r \cos \delta \right) \left(c^2_{12} c^2_{23} + s^2_{12} s^2_{23} s^2_{13} - 2 J_r \cos \delta \right) \Delta_{31}^2
\nonumber \\
A_{\mu \mu} &=& 4 s^2_{23} c^2_{13} \left( c^2_{23} + s^2_{23} s^2_{13} \right), 
\nonumber \\
s_{\mu} A_{\mu \mu} &=& 4 s^2_{23} c^2_{13} 
\left(c^2_{12} c^2_{23} + s^2_{12} s^2_{23} s^2_{13} - 2 J_r \cos \delta \right).
\label{Aeff-vac}
\end{eqnarray}
That is, $P(\nu _\mu \rightarrow \nu _\mu)$ in vacuum can be written in the effective two-flavor form.

Then, dividing the last line by the second, we obtain 
\begin{eqnarray}
s_{\mu} 
&=& c^2_{12} -
\frac{ \cos 2\theta_{12} \tan^2 \theta_{23} s^2_{13} + 2 \frac{J_r}{c^2_{23}} \cos \delta 
}{
1 + \tan^2 \theta_{23} s^2_{13} }.
\label{s-mu-vac}
\end{eqnarray}
We note that the matching between (\ref{Pmumu-vac2}) and (\ref{Pmumu-2flavor-2nd}) to order $\Delta_{21}^2$ is not possible with the current ansatz (\ref{P-alpha-alpha-vacuum-ansatz}) because the coefficient of $\Delta_{21}^2 \cos 2 \Delta_{31}$ term must be $A_{\mu \mu} s_{\mu}^2$, which does not mach with (\ref{Pmumu-vac2}).\footnote{
Introduction of the similar perturbative ansatz for $A_{\mu \mu}$ does not resolve this issue.
}
With $s_{\mu}$ in (\ref{s-mu-vac}), the effective $\Delta m^2_{\mu \mu} (= \Delta m^2_{31} - s_{\mu} \Delta m^2_{21})$ in vacuum is given to order $\epsilon s_{13}$ by the formula 
\begin{eqnarray}
\Delta m^2_{\mu \mu} 
&=& s^2_{12} \Delta m^2_{31} + c^2_{12} \Delta m^2_{32} + 2 \frac{J_r}{c^2_{23}} \cos \delta \Delta m^2_{21}, 
\label{Dm2-eff-mumu-vac}
\end{eqnarray}
which reproduces the expression of $\Delta m^2_{\mu \mu}$ in ref.~\cite{Nunokawa:2005nx}. 

A similar treatment with ansatz (\ref{P-alpha-alpha-vacuum-ansatz}) for $P(\nu _{e} \rightarrow \nu _{e})$ in vacuum gives the effective $\Delta m^2_{e e}$ without expanding by $s_{13}$ as 
\begin{eqnarray}
\Delta m^2_{ee} &=& c^2_{12} \Delta m^2_{31} + s^2_{12}  \Delta m^2_{32},  
\label{Dm2-eff-ee-vac}
\end{eqnarray}
again reproducing the formula for $\Delta m^2_{ee}$ in ref.~\cite{Nunokawa:2005nx}. In the rest of this paper, we will refer eqs.~(\ref{Dm2-eff-mumu-vac}) and (\ref{Dm2-eff-ee-vac}) as the NPZ formula for effective $\Delta m^2$. 

\section{Effective two-flavor form of survival probability in matter }
\label{sec:P-alpha-alpha-matt}

In matter, we don't know apriori whether the effective two-flavor form of the survival probability makes sense. Therefore, it is not obvious at all if there is such a concept as effective $\Delta m^2_{\alpha \alpha}(a) $ in matter. Fortunately, very recently, there was a progress in our understanding of this issue.

The authors of ref.~\cite{Minakata:2015gra} have shown to all orders in matter effect (with uniform density) as well as in $\theta_{13}$ that $P(\nu_{e} \rightarrow \nu_{e}: a)$ can be written in an effective two-flavor form 
\begin{eqnarray}
P(\nu_{e} \rightarrow \nu_{e}: a) &=& 
1 -  \sin^2 2 \tilde{\phi}~\sin^2 \frac{ (\lambda_{+} - \lambda_{-} ) L }{4E} 
\label{Pee-matter-SC}
\end{eqnarray}
to first order in their expansion parameter $\epsilon_r$, where $\tilde{\phi}$ is $\theta_{13}$ in matter and $\lambda_{\pm}$ denote the eigenvalues of the states which participate the 1-3 level crossing.\footnote{
In their framework, which is dubbed as the ``renormalized helio-perturbation theory'', they used a slightly different expansion parameter $\epsilon_r \equiv \Delta m^2_{21} / \Delta m^2_{ren}$, where $\Delta m^2_{ren} \equiv \Delta m^2_{31} - s^2_{12} \Delta m^2_{21}$, which is identical with $\Delta m^2_{ee}$ in vacuum, eq.~(\ref{Dm2-eff-ee-vac}). See ref.~\cite{Minakata:2015gra} for the explicit definitions of $\tilde{\phi}$, $\lambda_{\pm}$ etc.
}
This provides us an {\em existence proof} of the concept of the effective two-flavor form of the survival probability in matter. 

From (\ref{Pee-matter-SC}), $\Delta m^2_{ee} (a)$ in matter is given by (see \cite{Minakata:2015gra})
\begin{eqnarray}
\Delta m^2_{ee} (a) = 
\vert \lambda_{+} - \lambda_{-} \vert = 
\sqrt{ \left( \Delta m^2_{ren} - a \right)^2 + 4 s^2_{13} a \Delta m^2_{ren} } 
\label{Dm2ee-matter-SC}
\end{eqnarray}
where $a$ denotes the Wolfenstein matter potential \cite{Wolfenstein:1977ue} which in our convention depend on energy $E$ as 
\begin{eqnarray}
a & = & 2\sqrt{2} G_F N_e E \approx 1.52 \times 10^{-4} \left( \frac{Y_{e}~\rho}{\rm g.cm^{-3}} \right) \left( \frac{E}{\rm GeV} \right) {\rm eV}^2,  
\label{matt-potential}
\end{eqnarray}
where $G_F$ denotes the Fermi constant, $N_e$ the number density of electrons, $Y_e$ the electron fraction and $\rho$ is the density of matter. For simplicity and clarity we will work with the uniform matter density approximation in this paper. 

Then, the natural question is: Is the similar effective two-flavor form of the survival probability available in $\nu_{\mu}$ disappearance channel within the same framework? Unfortunately, the answer appears to be {\em No}.

One of the charming features of the framework developed in ref.~\cite{Minakata:2015gra} is that the oscillation probability takes the canonical form, the one with the same structure as in vacuum, of course with replacing the quantities $U_{\alpha i}$ and $\Delta_{ji}$ by the corresponding ones in matter. For the canonical or the vacuum-like structure of $P(\nu _\alpha \rightarrow \nu _\alpha)$, look at the first line in eq.~(\ref{P-alpha-alpha-vac}). Therefore, generally speaking, it contains the three terms with kinematic sine functions of the three differences of the eigenvalues, with exception of $P(\nu_{e} \rightarrow \nu_{e}: a)$ mentioned above. If one looks at the eigenvalue flow diagram as a function of the matter potential (figure 3 in \cite{Minakata:2015gra}) one would be convinced that there is no good reason to expect that the effective two-flavor form of the survival probability holds, except for the asymptotic regions $a \rightarrow \pm \infty$. After all, the system we are dealing with is the {\em three-flavor neutrino mixing } so that we must expect generically the genuine three-flavor structure.  

Then, the readers may ask: Is this the last word for answering the question ``Is there any sensible definition of effective two-flavor form of the survival probability in matter?''. Most probably the answer is {\em No}. Nonetheless, we will show in the rest of this paper that circumventing the conclusion in the last paragraph faces immediate difficulties. In a nutshell, we will show that the effective two-flavor form of the survival probability is possible in matter to order $\epsilon$ at least formally in both $\nu_{e}$ and $\nu_{\mu}$ disappearance channels. But, we show that the effective $\Delta m^2_{\mu \mu} (a)$ in matter cannot be regarded as a physically sensible quantity by being $L$ (baseline) dependent. On the other hand, $\Delta m^2_{ee} (a)$ does not suffer from the same disease.   

Our approach is that we limit ourselves to a simpler perturbative framework in which however the effect of matter to all orders is kept, because it is the key to the present discussion. It allows us to write the survival probability by simple analytic functions and the fact that each quantity has explicit form would allow us clearer understanding. 

\subsection{Ansatz for the effective two-flavor form of survival probability in matter}
\label{sec:ansatz}

With the above explicit example of $P(\nu_{e} \rightarrow \nu_{e}: a)$ in mind, we examine the similar ansatz as in vacuum for the effective two-flavor form of survival probability in matter. We postulate the same form of ansatz as in vacuum, which is assumed to be valid up to order $\epsilon$, but allowing more generic form of $\Delta m^2_{\alpha \alpha}$ ($\alpha= e, \mu, \tau$):
\begin{eqnarray} 
P(\nu _\alpha \rightarrow \nu _\alpha: a) &=&
C_{\alpha \alpha} (a) - A_{\alpha \alpha} (a) \sin ^2 \left(\frac{\Delta m^2_{\alpha \alpha} (a) L}{4E} \right), 
\label{P-alpha-alpha-matter} 
\\
\Delta m^2_{\alpha \alpha} (a) &=& 
\Delta m^2_{\alpha \alpha} (a)^{(0)} + \epsilon
\Delta m^2_{\alpha \alpha} (a)^{(1)}. 
\label{Dm2-eff-alpha-alpha}
\end{eqnarray}
The lessons we learned in the case in vacuum suggest that the restriction to order $\epsilon$ is necessary to keep the expression of the effective two-flavor probability sufficiently concise. Notice that the quantities $A_{\alpha \alpha}$, $C_{\alpha \alpha}$, and $\Delta m^2_{\alpha \alpha}$ in eq.~(\ref{P-alpha-alpha-matter}) depend not only on the mixing parameters but also on the matter potential $a$, as explicitly indicated in (\ref{P-alpha-alpha-matter}).

We then follow the procedure in section~\ref{sec:2flavor-vac} to determine the form of $\Delta m^2_{\alpha \alpha} (a)^{(0)}$ and $\Delta m^2_{\alpha \alpha} (a)^{(1)}$. As was done in the previous section we occasionally use a concise notation 
$\Delta_{\alpha \alpha} (a) \equiv \frac{ \Delta m^2_{\alpha \alpha} (a) L }{ 4E }$. The effective two-flavor form of $P(\nu_{\alpha} \rightarrow \nu_{\alpha})$, eq.~(\ref{P-alpha-alpha-matter}), can be expanded in terms of $\epsilon$
\begin{eqnarray} 
P(\nu_\alpha \rightarrow \nu _\alpha: a) &=&
C_{\alpha \alpha} (a) - A_{\alpha \alpha} (a)
\left( \sin^2 \Delta_{\alpha \alpha}^{(0)} (a) + \epsilon \Delta_{\alpha \alpha}^{(1)} (a) \sin 2 \Delta_{\alpha \alpha}^{(0)} (a) \right) 
+ \mathcal{O} (\epsilon^2).
\label{Palpha-alpha-2flavor3}
\end{eqnarray}
Therefore, if the expressions of the survival probabilities take the form (\ref{Palpha-alpha-2flavor3}), then they can be written as the effective two flavor forms which are valid to order $\epsilon$.

\subsection{Perturbative framework to compute the oscillation probabilities}
\label{sec:framework}

We use the $\sqrt{\epsilon}$ perturbation theory formulated in ref.~\cite{Asano:2011nj} to derive the suitable expressions of the survival probabilities in matter. In this framework the oscillation probabilities are computed to a certain desired order of the small expansion parameter $\epsilon \equiv \Delta m^2_{21} / \Delta m^2_{31} \simeq 0.03$ assuming $s_{13} \sim \sqrt{\epsilon} $. Notice that the measured value of $\theta_{13}$ is $s_{13} = 0.147$, the central value of the largest statistics measurement \cite{An:2016ses}, so that $s^2_{13} = 0.021 \sim \epsilon$. We use the survival probabilities computed to second order in $\epsilon$, which means to order $\epsilon s^2_{13}$ and $s^4_{13}$. Inclusion of these higher-order corrections implies to go beyond the Cervela et al. formula \cite{Cervera:2000kp}. We will see that it is necessary to keep the former higher-order term to recover the NPZ formula in the vacuum limit.

While we expand the oscillation probabilities in terms of $\epsilon$ and $s_{13}$, we keep the matter effect to all orders. It is the key to our discussion, and furthermore keeping all-order effect of matter may widen the possibility of application of this discussion to various experimental setups of the long-baseline (LBL) accelerator neutrino experiments considered in the literature. The parameter which measures relative importance of the matter effect to the vacuum one is given by 
\begin{eqnarray} 
r_{A} &\equiv& \frac{ a }{ \Delta m^2_{31} } = 
\frac{ 2\sqrt{2} G_F Y_{e} \rho} { \Delta m^2_{31} m_{N} } E 
\nonumber \\
&=&
0.89 
\left(\frac{|\Delta m^2_{31}|}{2.4 \times 10^{-3}\mbox{eV}^2}\right)^{-1}
\left(\frac{\rho}{2.8 \text{g/cm}^3}\right) \left(\frac{E}{10~\mbox{GeV}}\right), 
\label{rA-def-value}
\end{eqnarray}
where $m_{N}$ denotes the unified atomic mass unit, and we assume $Y_{e}=0.5$ in this paper. 
$r_{A}$ appears frequently in the expressions of the oscillation probabilities, as will be seen below. Notice that the ratio $r_{A}$ of matter to vacuum effects can be sizeable for neutrino energies of $\sim 10$ GeV in the LBL experiments. It should also be noticed that $r_{A}$ depends linearly on neutrino energy $E$. In what follows, we use the formulas of the probabilities given in \cite{Asano:2011nj} without explanation, leaving the derivation to the reference.

\section{Effective $\Delta m^2_{ee}$ in matter }
\label{sec:Dm2-ee-matt}

Given the perturbative expressions of the survival probabilities we can derive the effective $\Delta m^2_{\alpha \alpha} (a)$ in matter by using the matching condition with (\ref{Palpha-alpha-2flavor3}), in the similar way as done in section~\ref{sec:2flavor-vac}. In addition to the abbreviated notation $\Delta_{j i} \equiv \frac{\Delta m^2_{j i} L }{4E}$ introduced in (\ref{Delta-ji-def}), we use the notation 
\begin{eqnarray}
\Delta_{a} \equiv \frac{a L }{4E} = r_{A} \Delta_{31}
\label{Delta-a-def}
\end{eqnarray}
with the matter potential $a$ defined in (\ref{matt-potential}) to simplify the expressions of the oscillation probabilities. Notice that, unlike $r_{A}$, $\Delta_{a}$ is energy independent, but $\Delta_{a}$ depends on the baseline $L$, $\Delta_{a} \propto L$. 

\subsection{Effective two-flavor form of $P(\nu_{e} \rightarrow \nu_{e})$ and $\Delta m^2_{ee}$ in matter}
\label{sec:Pee-matt}

We first discuss the $\nu_{e}$ survival probability $P(\nu_{e} \rightarrow \nu_{e})$ in matter. As we remarked in section~\ref{sec:framework}
we need to go to second order in $\epsilon$:\footnote{
Here is a comment on behaviour of $P(\nu_{e} \rightarrow \nu_{e}: a)$ in region of energy for $r_{A} \simeq 1$. Though it may look like that $P(\nu_{e} \rightarrow \nu_{e}: a)$ is singular in $r_{A} \rightarrow 1$ limit, it is not true. The apparent singularity cancels. But, it is not the end of the story. Despite no singularity at $r_{A}=1$, the perturbative expressions of the oscillation probabilities in region of $r_{A}$ close to 1 display the problem of inaccuracy. The cause of the problem is due to the fact that we are expanding the probability by $s_{13}$, by which we miss the effect of resonance enhancement of flavor oscillation. If fact, one can observe the improvement of the accuracy at around $r_{A}$ close to 1 by including $s^4_{13}$ terms. See figure 3 of ref.~\cite{Asano:2011nj}. 
}
\begin{eqnarray}
P(\nu_{e} \rightarrow \nu_{e}: a) &=& 
1 - 4 s^2_{13} \frac{ 1 }{ (1 - r_{A})^2 } 
\sin^2 \left[ (1 - r_{A}) \Delta_{31} \right] 
\nonumber \\
&+& 
4 \left[ 
s^4_{13} \frac{ (1 + r_{A})^2 }{ (1 - r_{A})^4 } 
- 2 s^2_{12} s^2_{13} \frac{ \epsilon r_{A} }{ (1 - r_{A})^3 } 
\right] 
\sin^2 \left[ (1 - r_{A}) \Delta_{31} \right] 
\nonumber \\
&-& 
4 \left[ 
2 s^4_{13} \frac{ r_{A} }{ (1 - r_{A})^3 } 
- s^2_{12} s^2_{13} \frac{ \epsilon }{ (1 - r_{A})^2 } 
\right] 
\Delta_{31} \sin \left[ 2 (1 - r_{A}) \Delta_{31} \right] 
\nonumber \\
&-&
4 c^2_{12} s^2_{12} 
\left( \frac{ \epsilon }{ r_{A} } \right)^2 
\sin^2 \Delta_{a}.
\label{Pee-matter}
\end{eqnarray}
The leading order depletion term in $P(\nu_{e} \rightarrow \nu_{e}: a)$ in (\ref{Pee-matter}) is of order $\epsilon$, and the remaining terms (second to fourth lines) are of order $\epsilon^2$. $\bar{\nu}_{e}$ survival probability can be discussed just by flipping the sign of the matter potential $a$.

We notice in eq.~(\ref{Pee-matter}) that even in the two flavor limit, $\epsilon \rightarrow 0$ and $s_{13} \rightarrow 0$, the effective $\Delta m^2$ is modified from $\Delta m^2_{31}$ to $\Delta m^2_{ee} (a)^{(0)} = (1 - r_{A}) \Delta m^2_{31}$ due to the strong, order unity, matter effect in the $\nu_{e}$ channel. Notice that in view of eq.~(\ref{rA-def-value}) the change can be sizeable at energies $E \gsim$ a few GeV. Thus, the effective $\Delta m^2_{ee} (a)$ in matter, and generically $\Delta m^2_{\alpha \alpha} (a)$ as we will see later, inevitably become dynamical quantities, which depend on neutrino energy $E$.

The matching between (\ref{Pee-matter}) and the two-flavor form in (\ref{Palpha-alpha-2flavor3}) can be achieved as follows:  
\begin{eqnarray}
C_{ee} (a) &=& 1 - 4 c^2_{12} s^2_{12} 
\left( \frac{ \epsilon }{ r_{A} } \right)^2 
\sin^2 \Delta_{a}, 
\nonumber \\
A_{ee} (a) &=& 
\frac{ 4 s^2_{13} }{ (1 - r_{A})^2 } 
\left[ 1 
+ 2 s^2_{12} \frac{  \epsilon r_{A} }{ (1 - r_{A}) } 
- s^2_{13} \frac{ (1 + r_{A})^2 }{ (1 - r_{A})^2 } 
\right], 
\nonumber \\
\epsilon A_{ee} (a) \Delta m^2_{ee} (a)^{(1)} &=& 
\frac{ 4 s^2_{13} }{ (1 - r_{A})^2 } 
\left[
2 s^2_{13} \frac{ r_{A} }{ (1 - r_{A}) } 
- s^2_{12}  \epsilon
\right] \Delta m^2_{31}.
\label{coeff-ee-matter}
\end{eqnarray}
It is remarkable to see that all the terms in (\ref{Pee-matter}) including $\mathcal{O} (\epsilon^2)$ terms can be organized into the effective two-flavor form in (\ref{P-alpha-alpha-matter}). Using the second and the fourth lines of (\ref{coeff-ee-matter}) we obtain to first order in $\epsilon$: 
\begin{eqnarray}
\epsilon \Delta m^2_{ee} (a)^{(1)}
=
\frac{\epsilon A_{ee} (a) \Delta m^2_{ee} (a)^{(1)} }{ A_{ee} (a) } 
= \left[
2 s^2_{13} \frac{ r_{A} }{ (1 - r_{A}) } 
- s^2_{12} \epsilon
\right]
\Delta m^2_{31} 
\label{Dm2-1st}
\end{eqnarray}
where we have kept terms up to order $\epsilon$ in the second line in (\ref{Dm2-1st}). Thus, the effective $\Delta m^2$ in matter in the $\nu _e \rightarrow \nu _e$ channel is given as $\Delta m^2_{ee} (a) = \Delta m^2_{ee} \vert^{(0)} + \epsilon \Delta m^2_{ee} (a)^{(1)}$, 
\begin{eqnarray}
\Delta m^2_{ee} (a) &=& 
(1 - r_{A}) \Delta m^2_{31} + 
\left[ 2 s^2_{13} \frac{ r_{A} }{ (1 - r_{A}) } - \epsilon s^2_{12} 
\right] \Delta m^2_{31},
\nonumber \\ &=& 
(1 - r_{A}) \Delta m^2_{ee} (0) +  
r_{A} \left[ \frac{ 2 s^2_{13} }{ (1 - r_{A}) } - \epsilon s^2_{12} \right] \Delta m^2_{31}, 
\label{Dm2-eff-ee-matter}
\end{eqnarray}
which obviously reduces to the NPZ formula $\Delta m^2_{ee} \vert_{ \text {vac} } =\Delta m^2_{ee} (0) = c^2_{12} \Delta m^2_{31} + s^2_{12} \Delta m^2_{32}$ in the vacuum limit. 

Thus, we have learned that $\nu_{e}$ (and $\bar{\nu}_{e}$) survival probability in matter can be casted into the effective two-flavor form (\ref{P-alpha-alpha-matter}) in a way parallel to that in vacuum. But, the nature of the effective $\Delta m^2_{ee} (a)$ is qualitatively changed in matter: It becomes a dynamical quantity which depends on energy, not just a combination of fundamental parameters as it is in vacuum. It is inevitable once we recognize that the leading-order effective $\Delta m^2_{ee}$ in matter is given by $\Delta m^2_{ee} (a)^{(0)} = (1 - r_{A}) \Delta m^2_{31}$ in the two-flavor limit. 

One may argue that the expression of $\Delta m^2_{ee} (a)$ in eq.~(\ref{Dm2-eff-ee-matter}) does not make sense because it is singular at $r_{A} \rightarrow 1$ limit. It might sound a very relevant point because the survival probability itself is singularity free, as mentioned in the footnote 4. But, we argue that the singularity of $\Delta m^2_{ee} (a)$ at $r_{A} =1$ is very likely to be superficial. Let us go back to $\Delta m^2_{ee} (a)$ in eq.~(\ref{Dm2ee-matter-SC}) which is obtained by using the renormalized helio-perturbation theory \cite{Minakata:2015gra} with all order effect of $\theta_{13}$. It is perfectly finite in the limit $r_{A} \rightarrow 1$. One can easily show that by expanding $\Delta m^2_{ee} (a)$ in (\ref{Dm2ee-matter-SC}) by $s_{13}$ one reproduces the result in (\ref{Dm2-eff-ee-matter}).\footnote{
This exercise has first been suggested to the author by Stephen Parke. 
}
It means that the singularity in $\Delta m^2_{ee} (a)$ at $r_{A} = 1$ is an artifact of the expansion around $s_{13}=0$. In fact, one can easily convince oneself that the expansion of the eigenvalues in terms of $s_{13}$ is actually an expansion in terms of $s^2_{13}\frac{r_{A}}{ (1-r_{A})^2 }$. 

\subsection{Energy dependence of $\Delta m^2_{ee} (a)$} 
\label{sec:E-dep-Dm2}

\begin{figure}
\begin{center}
\vspace{3mm}
\includegraphics[width=0.48\textwidth]{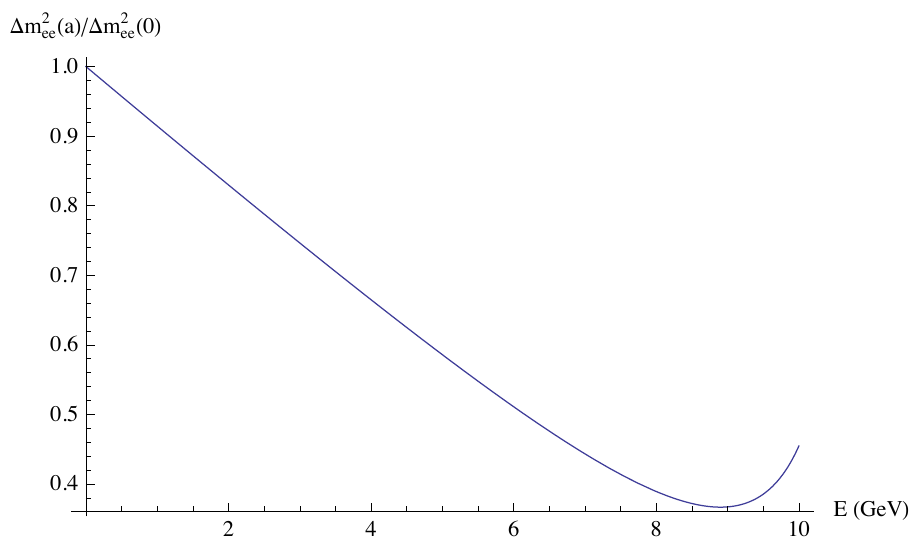}
\includegraphics[width=0.48\textwidth]{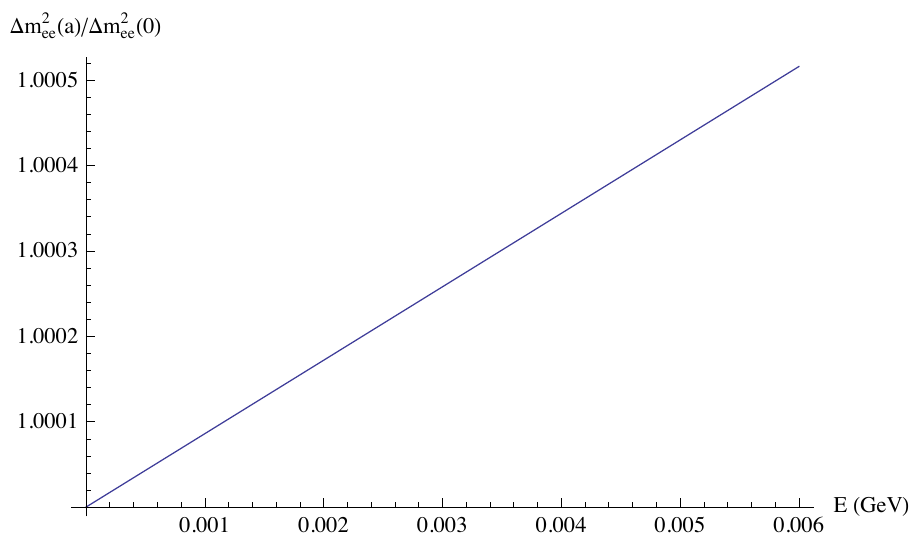}
\end{center}
\caption{
In the left panel, plotted is the ratio $\Delta m^2_{ee} (a) / \Delta m^2_{ee} (0)$ as a function of $E$ in units of GeV. The right panel is to magnify the low energy region of $\Delta m^2_{ee} (a) / \Delta m^2_{ee} (0)$ for anti-neutrinos, 
showing that the matter effect in $\Delta m^2_{ee} (a)$ is tiny, at a level of $\sim \text{a few} \times 10^{-4}$ in MeV energy region. The mixing parameters and the matter density that we used are: 
$\Delta m^2_{31} = 2.4 \times 10^{-3}$ eV$^2$, $\Delta m^2_{21} = 7.5 \times 10^{-5}$ eV$^2$, $\sin^2 \theta_{13} = 0.022$, $\sin^2 \theta_{12} = 0.30$, and $\rho=2.8~\text{g/cm}^3$. 
}
\label{fig:Dm2-ratio-ee}
\end{figure}

In figure~\ref{fig:Dm2-ratio-ee}, the ratio $\Delta m^2_{ee} (a) / \Delta m^2_{ee} (0)$ is plotted as a function of neutrino energy $E$ in units of GeV. The left panel of Fig.~\ref{fig:Dm2-ratio-ee} indicates that $\Delta m^2_{ee} (a)$ decreases linearly with $E$ in a good approximation, the behaviour due to the leading order term $\Delta m^2_{ee} (a)^{(0)} = (1 - r_{A}) \Delta m^2_{31}$. Our expression of $\Delta m^2_{ee} (a)$ cannot be trusted beyond $E \simeq 7$ GeV because the turn over behaviour seen in figure~\ref{fig:Dm2-ratio-ee} starting at the energy signals approach to the resonance enhancement at $E \simeq 11$ GeV. An estimation of the resonance width via the conventional way yields the results $\pm 3.3$ GeV around the resonance, inside which our perturbation theory breaks down. The estimated width is consistent with what we see in figure~\ref{fig:Dm2-ratio-ee}. The deviation from the linearity below that energy represents the effect of three-flavor correction, the second term in the last line in (\ref{Dm2-eff-ee-matter}), and its smallness indicates that this effect is small, and it is nicely accommodated into the effective two-flavor $\Delta m^2_{ee} (a)$. 

The right panel in Fig.~\ref{fig:Dm2-ratio-ee} shows $\Delta m^2_{ee} (a)$ for the antineutrino channel at low energies relevant for reactor electron antineutrinos. We see that the matter effect is extremely small, $0.05\%$ even at $E=6$ MeV, which justifies the commonly used vacuum approximation for $\Delta m^2_{ee}$ for reactor neutrino analyses \cite{An:2016ses,RENO:2015ksa}. 

\subsection{Energy dependence of the minimum of $P(\nu_{e} \rightarrow \nu_{e})$} 
\label{sec:E-dep-Peemin}

In section~\ref{sec:Pee-matt}, the formula for the effective $\Delta m^2_{ee} (a)$ in matter was derived in an analytic way, eq.~(\ref{Dm2-eff-ee-matter}). The question we want to address in this section is to what extent the energy dependent $\Delta m^2_{ee} (a)$ is sufficient to describe the behaviour of $\nu_{e}$ disappearance probability at around $E=E_{min}$, the highest-energy  minimum of $P(\nu_{e} \rightarrow \nu_{e}: a)$. For this purpose, we construct a simple model of $P(\nu_{e} \rightarrow \nu_{e}: a)$ in which the matter (therefore energy) dependence exists only in $\Delta m^2_{ee} (a)$: 

\vspace{2mm}
\noindent
{\bf Simple model}: We ignore the energy dependence of $C_{ee} (a)$ and $A_{ee} (a)$ in the effective two-flavor form eq.~(\ref{P-alpha-alpha-matter}) of $P(\nu_{e} \rightarrow \nu_{e}: a)$, while keeping the energy dependence in $\Delta m^2_{ee} (a)$.

\vspace{2mm}

\noindent
The spirit of the model is that the energy dependent $\Delta m^2_{ee} (a)$ plays a dominant role in describing the behaviour of $P(\nu_{e} \rightarrow \nu_{e}: a)$ at around $E=E_{min}$. We want to test this simple model to know to what extent the spirit is shared by the actual $P(\nu_{e} \rightarrow \nu_{e}: a)$ in matter. 

\begin{figure}
\begin{center}
\vspace{-4.4mm}
\includegraphics[width=0.48\textwidth]{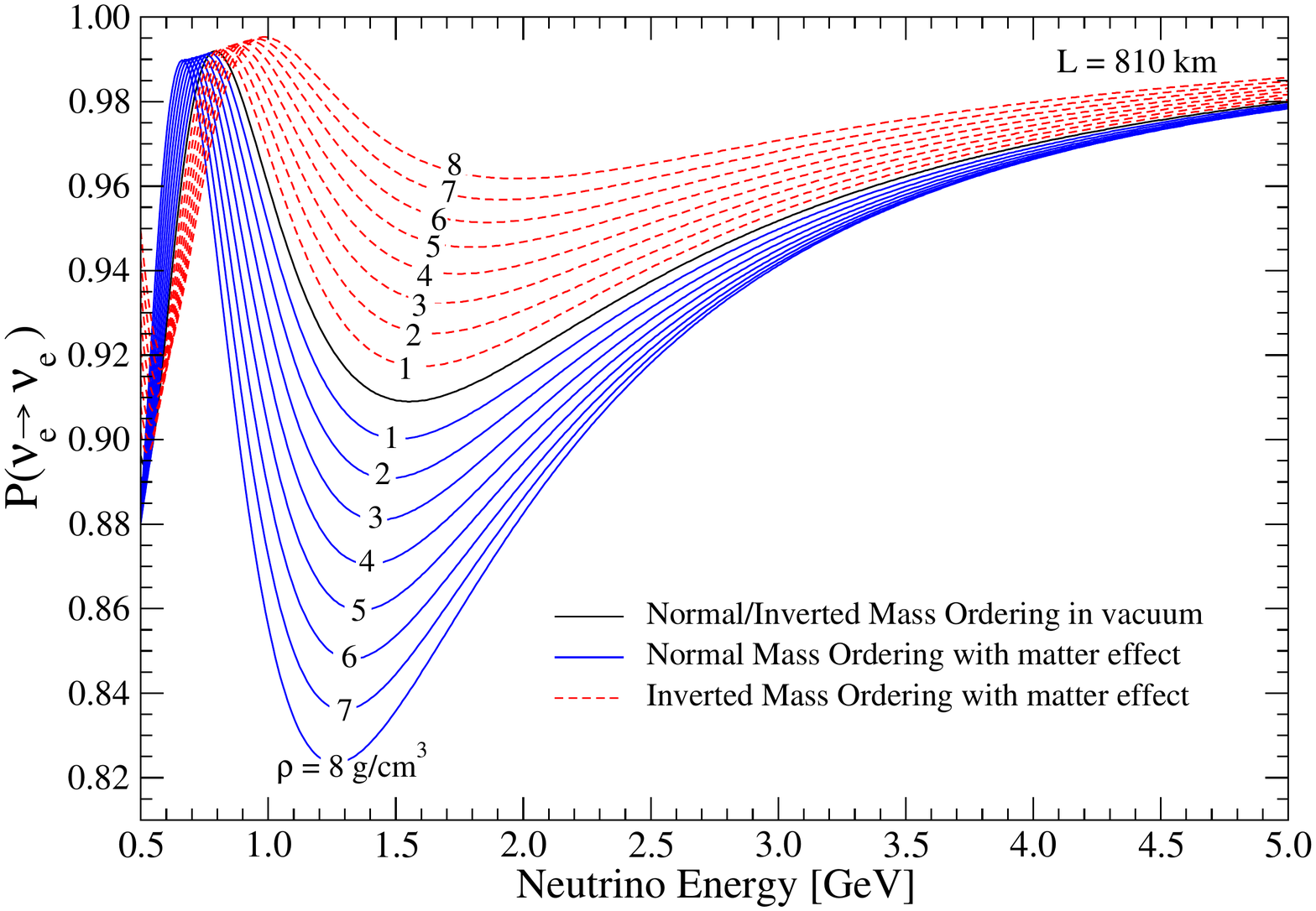}
\includegraphics[width=0.48\textwidth]{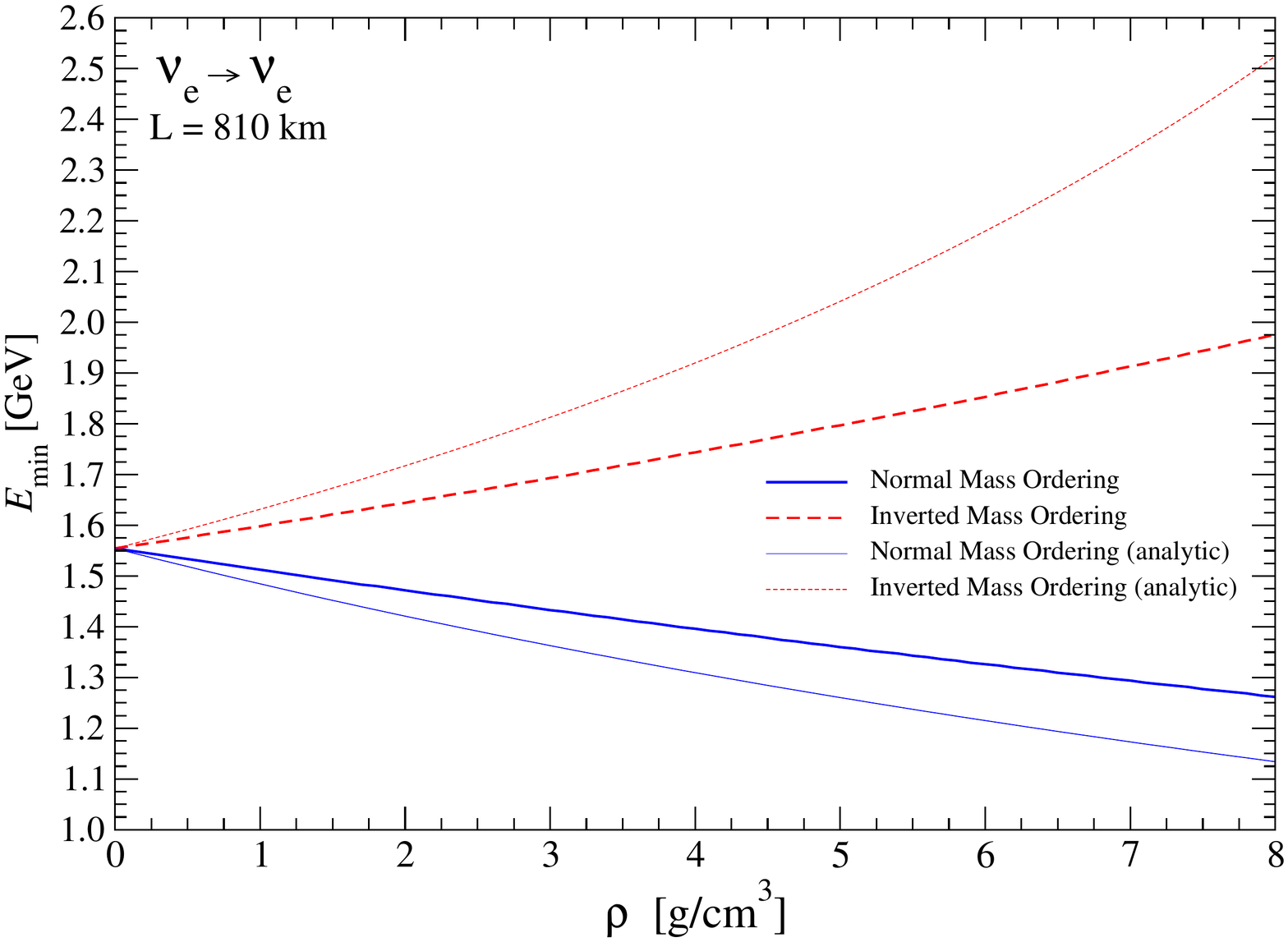}
\end{center}
\vspace{-6mm}
\caption{ The left panel: The survival probability $P(\nu_{e} \rightarrow \nu_{e}: a)$ is plotted as a function of neutrino energy $E$ obtained by numerically solving the neutrino evolution equation for various values of matter density between $\rho = 0$ and $\rho = 8$ $\frac{ \text{g} }{ \text{cm}^3}$. The blue-solid and red-dashed lines are for the normal and inverted mass orderings, respectively. 
The right panel: The highest-energy solution $E_{min}$ of the equation $\frac{ d }{d E} P(\nu_{e} \rightarrow \nu_{e}: a) = 0$ is plotted as a function of $\rho$ in units of $\frac{ \text{g} }{ \text{cm}^3}$. $E_{min}$ is with use of the same color line symbols as in the left panel. Also plotted are the solution of eq.~(\ref{Emin-eq}) obtained in the simple model described in the text.
The mixing parameters used are: 
$\Delta m^2_{ee} = 2.4 \times 10^{-3}$ eV$^2$, $\Delta m^2_{21} = 7.54 \times 10^{-5}$ eV$^2$, $\sin^2 \theta_{12} = 0.31$, and $\sin^2 2\theta_{13} = 0.089$. 
}
\label{fig:Pee}
\end{figure}

In figure~\ref{fig:Pee}, in the left panel, plotted is the survival probability $P(\nu_{e} \rightarrow \nu_{e}: a)$ as a function of neutrino energy $E$ obtained by solving exactly (within numerical precision) the neutrino evolution equation for various values of matter density between $\rho = 0$ and $\rho = 8$ $\frac{ \text{g} }{ \text{cm}^3}$ to vary the strength of the matter effect.
%
The blue-solid and red-dashed lines are for the normal and inverted mass orderings, respectively. In mid between the blue and red colored lines there is a black solid line which corresponds to $P(\nu_{e} \rightarrow \nu_{e})$ in vacuum. 
In the right panel in figure~\ref{fig:Pee}, plotted with the same line symbols as in the left panel is the highest-energy solution $E_{min}$ of the equation $\frac{ d }{d E} P(\nu_{e} \rightarrow \nu_{e}: a) = 0$ as a function of $\rho$ in units of $\frac{ \text{g} }{ \text{cm}^3}$. $E_{min}$ corresponds to so called the dip energy at the first minimum of $P(\nu_{e} \rightarrow \nu_{e}: a)$. The thin blue-solid and red-dotted lines are the solution of $\frac{ d }{d E} P(\nu_{e} \rightarrow \nu_{e}: a) = 0$ of the simple model for the normal and inverted mass orderings, respectively. 

We now try to understand qualitatively figure~\ref{fig:Pee}, and compare $E_{min}$ predicted by the simple model to the one obtained by using the numerically computed survival probability. The solution of $\frac{ d }{d E} P(\nu_{e} \rightarrow \nu_{e}: a) = 0$ in the simple model is given by 
\begin{eqnarray}
\frac{ \Delta m^2_{ee} (E) L }{ 2 E } = \pm \pi, 
\label{E-min}
\end{eqnarray}
where the sign $\pm$ corresponds to the normal and inverted mass orderings, respectively. 
To simplify the expression we use the notations 
\begin{eqnarray}
r_{A} &\equiv& \pm A E, 
\hspace{10mm}
A \equiv \frac{ 2\sqrt{2} G_F Y_{e} \rho} { \vert \Delta m^2_{31} \vert m_{N} }, 
\hspace{10mm} 
E_{vom} \equiv \frac{ \vert \Delta m^2_{31} \vert L }{ 2 \pi }.
\label{notations}
\end{eqnarray}
Then, by using (\ref{Dm2-eff-ee-matter}) the $E_{min}$-determining equation (\ref{E-min}) becomes 
\begin{eqnarray}
1 \mp AE 
\pm \left[ 2 s^2_{13} \frac{ AE }{ 1 \mp AE } - \epsilon s^2_{12} \right] = 
\frac{ E }{ E_{vom} }.
\label{Emin-eq}
\end{eqnarray}
It is a quadratic equation for $E$ with an obvious solution that is not written here. The solution of (\ref{Emin-eq}) is plotted by the thin-solid and dotted lines in the right panel of figure~\ref{fig:Pee}. 

The qualitative behaviour of the solution of (\ref{Emin-eq}) can be understood by a perturbative solution of (\ref{Emin-eq}) with the small parameters $\epsilon$ and $s^2_{13} \sim \epsilon$. To first order in $\epsilon$ it reads 
\begin{eqnarray}
E_{min} = 
\frac{ E_{vom} }{ 1 \pm A E_{vom}  }
\left[ 1 \pm \left( 2 s^2_{13} A E_{vom}  - \epsilon s^2_{12} \right) \right]. 
\label{Emin-sol}
\end{eqnarray}
Noticing the value of $A$, 
\begin{eqnarray} 
A &=& 
\frac{ 2\sqrt{2} G_F Y_{e} \rho} { \vert \Delta m^2_{31} \vert m_{N} } = 
0.032 
\left(\frac{|\Delta m^2_{31}|}{2.4 \times 10^{-3}\mbox{eV}^2}\right)^{-1}
\left(\frac{\rho}{1 \text{g/cm}^3}\right) \mbox{GeV}^{-1}, 
\label{A-def-value}
\end{eqnarray}
which is small for $\rho \lsim 3~\text{g/cm}^3$, an approximately linear $\rho$ dependence $E_{min} \approx E_{vom} \left( 1 \mp A E_{vom} \right)$ is expected. But, in region of $\rho \gsim 6~\text{g/cm}^3$ a visible nonlinearity is expected in particular in the case of inverted mass ordering. They are in good agreement with the simple model prediction plotted by the thin-solid and dotted lines in the right panel of figure~\ref{fig:Pee}. It confirms that the perturbative solution (\ref{Emin-sol}) captures the main feature of the simple model. We note, however, that the agreement between the simple model prediction and the numerically computed $E_{min}$ (blue-solid and red-dashed lines) is rather poor as seen in the same figure. 

Thus, despite qualitative consistency exists to certain extent, we see that the simple model fails to explain the quantitative features of $\rho$ dependence of the first minimum of $P(\nu_{e} \rightarrow \nu_{e}: a)$. It indicates that the energy dependent $\Delta m^2_{ee} (a)$ is not sufficient to describe the behaviour of $\nu_{e}$ disappearance probability at around its first minimum. That is, the matter effect brings the energy dependences into the coefficients $C_{ee}$ and $A_{ee}$ in (\ref{P-alpha-alpha-matter}) as strongly as to modify $\Delta m^2_{ee} (a)$. Therefore, though the perfectly consistent effective two-flavor approximation exists for $\nu_{e}$ survival probability in matter, its quantitative behaviour at around the highest-energy minimum cannot be described solely by the energy-dependent $\Delta m^2_{ee} (a)$. This is in contrast to the situation in vacuum that introduction of $\Delta m^2_{ee}$ allows to describe the result of precision measurement of $P(\nu_{e} \rightarrow \nu_{e})$ in reactor experiments very well \cite{An:2016ses,RENO:2015ksa}.

\section{Effective $\Delta m^2_{\mu \mu}$ in matter }
\label{sec:Dm2-mumu-matt}

\subsection{Effective two-flavor form of $P(\nu_{\mu} \rightarrow \nu_{\mu})$ and $\Delta m^2_{\mu \mu}$ in matter}
\label{sec:Pmumu-matt}

We now discuss $\Delta m^2_{\mu \mu}$ in matter. Here, we need the survival probability $P(\nu _\mu \rightarrow \nu _\mu; a)$ only up to second order in $s_{13}$ and first order in $\epsilon \equiv \frac{\Delta m^2_{21}}{\Delta m^2_{31}}$, because the leading order term is of order unity. These terms were calculated previously by many authors, see e.g., \cite{Cervera:2000kp,Akhmedov:2004ny,Minakata:2004pg}. It can be written as the effective two flavor form (\ref{Palpha-alpha-2flavor3}) with the coefficients 
\begin{eqnarray}
C_{\mu \mu} (a) &=& 
1 - 4 \biggl[ 
s^2_{23} s^2_{13}  \left(\frac{1}{1 - r_{A}}\right)^2  - 2 \epsilon J_r \cos \delta \frac{1}{r_{A} (1 - r_{A})} \biggr] 
\sin ^2 \Delta_{a}, 
\nonumber \\
A_{\mu \mu} (a) &=& 
4 
\left[
c^2_{23} s^2_{23} - s^2_{23} s^2_{13} \left(\frac{1}{1 - r_{A}}\right)^2 
\left( \cos 2\theta_{23} + 2 s^2_{23} \sin ^2 \Delta_{a} \right) 
 \right. 
  \nonumber \\ 
  && \left. 
   \hspace{18mm} 
+ 2 \epsilon J_r \cos \delta \frac{1}{r_{A} (1 - r_{A})} 
\left( 
\cos 2\theta_{23} r_{A}^2 + 2 s^2_{23} \sin^2\Delta_{a} 
\right) 
\right], 
%
%
\nonumber \\
\epsilon \Delta m^2_{\mu \mu} (a)^{(1)} A_{\mu \mu} (a) &=& 
2 s^2_{23} 
\left[
2 c^2_{23} \biggl\{ 
s^2_{13} \left(\frac{r_{A}}{1 - r_{A}} \right) 
- \epsilon c^2_{12}  
\biggr\}
\right. 
\nonumber \\ 
&& \left. 
\hspace{-6mm} 
- \biggl\{ s^2_{23} s^2_{13} 
\left(\frac{1}{1 - r_{A}}\right)^2  
- 2 \epsilon J_r \cos \delta \frac{1}{r_{A} (1 - r_{A})} 
\biggr\} \frac{ \sin 2\Delta_{a} }{ \Delta_{31} }
\right] 
\Delta m^2_{31}, 
%
%
\label{matching2}
\end{eqnarray}
where $J_r \equiv c_{12} s_{12} c_{23} s_{23} s_{13}$. 

Using the last two equations in (\ref{matching2}), the first order correction term in the effective $\Delta m^2_{\mu \mu} (a)$ can be calculated, to order $s^2_{13} \sim \epsilon$ and $\epsilon s_{13}$, as 
\begin{eqnarray}
&& \epsilon \Delta m^2_{\mu \mu} (a)^{(1)} = 
\frac{ \epsilon \Delta m^2_{\mu \mu} (a)^{(1)} A_{\mu \mu} (a) }{ A_{\mu \mu} (a) }
\nonumber \\ 
&=&
\left[
- \epsilon c^2_{12} 
+ s^2_{13} \left(\frac{r_{A}}{1 - r_{A}} \right) 
- \biggl\{ \frac{ 1 }{ 2 } s^2_{13} \tan^2 \theta_{23} 
\frac{1}{( 1 - r_{A} )^2 } 
- \epsilon \frac{ J_r \cos \delta }{ c^2_{23}  } \frac{1}{r_{A} (1 - r_{A})} 
\biggr\} \frac{ \sin 2\Delta_{a} }{ \Delta_{31} } 
\right] \Delta m^2_{31}.
\nonumber \\ 
\label{Dm2-mumu-1st-again}
\end{eqnarray}
Then, finally, $\Delta m^2_{\mu \mu} (a)= \Delta m^2_{\mu \mu} (a)^{(0)} + \epsilon \Delta m^2_{\mu \mu} (a)^{(1)}$ can be obtained as 
\begin{eqnarray}
&& \Delta m^2_{\mu \mu} (a) = 
s^2_{12} \Delta m^2_{31} + c^2_{12} \Delta m^2_{32} 
\nonumber \\
&+& \left[
s^2_{13} \left\{ \frac{r_{A}}{1 - r_{A}} 
- \tan^2 \theta_{23} 
\left(\frac{1}{1 - r_{A}}\right)^2 \frac{ \sin 2\Delta_{a} }{ 2 \Delta_{31} } 
\right\} 
+ 2 \frac{ \epsilon J_r \cos \delta }{ c^2_{23}  } \frac{1}{ (1 - r_{A})} 
\frac{ \sin 2\Delta_{a} }{ 2 \Delta_{a} } 
\right] \Delta m^2_{31}.
\nonumber \\
\label{Dm2-eff-mumu-matt}
\end{eqnarray}

In the vacuum limit, noticing that $\epsilon \Delta m^2_{\mu \mu} (a)^{(1)} \rightarrow 
\left( - c^2_{12} + 2 \frac{ J_r }{ c^2_{23}  } \cos \delta \right) \Delta m^2_{21}$ as $a \rightarrow 0$, we obtain 
\begin{eqnarray}
\Delta m^2_{\mu \mu} (0) 
&=& s^2_{12} \Delta m^2_{31} + c^2_{12} \Delta m^2_{32} 
+ 2 \frac{J_r}{c^2_{23}} \cos \delta~\Delta m^2_{21}, 
\label{Dm2-eff-mumu-vac2}
\end{eqnarray}
which again reproduces the NPZ formula for $\Delta m^2_{\mu \mu}$ in vacuum. 

Now, we have to address the conceptual issue about the result of $\Delta m^2_{\mu \mu} (a)$ in (\ref{Dm2-eff-mumu-matt}). Though it depends on energy through $r_{A} \propto E$ we do not think it a problem. See the discussion in the previous section. However, there is a problem of $L$-dependence of $\Delta m^2_{\mu \mu} (a)$. 
Notice that $\Delta_{a} \equiv aL/ 4E$ is $L$-dependent and is $E$ independent. Therefore, the last two terms of $\Delta m^2_{\mu \mu} (a)$ in (\ref{Dm2-eff-mumu-matt}) have a peculiar dependence on baseline length $L$.\footnote{
In short baseline, or in low-density medium, $\Delta_{a} \ll 1$, the $L$ dependence in (\ref{Dm2-eff-mumu-matt}) goes away because 
\begin{eqnarray}
\frac{ \sin 2\Delta_{a} }{ 2 \Delta_{31} } \approx 
\frac{ \Delta_{a} }{ \Delta_{31} } = r_{A}, 
\hspace{10mm}
\frac{ \sin 2\Delta_{a} }{ 2 \Delta_{a} } \approx 1.
\end{eqnarray}
But, this is just very special cases of possible experimental setups. 
}
Because of the $L$-dependence of $\Delta m^2_{\mu \mu} (a)$ in (\ref{Dm2-eff-mumu-matt}), unfortunately, we cannot consider it as the sensible quantity as the effective parameter which describes the physics of $\nu_{\mu}$ survival probability in matter.\footnote{
Some examples of $L$-dependent (actually $L/E$-dependent) effective $\Delta m^2_{ee}$ in vacuum are discussed recently with the critical comments \cite{Parke:2016joa}.
}

\begin{figure}
\begin{center}
\includegraphics[width=0.6\textwidth]{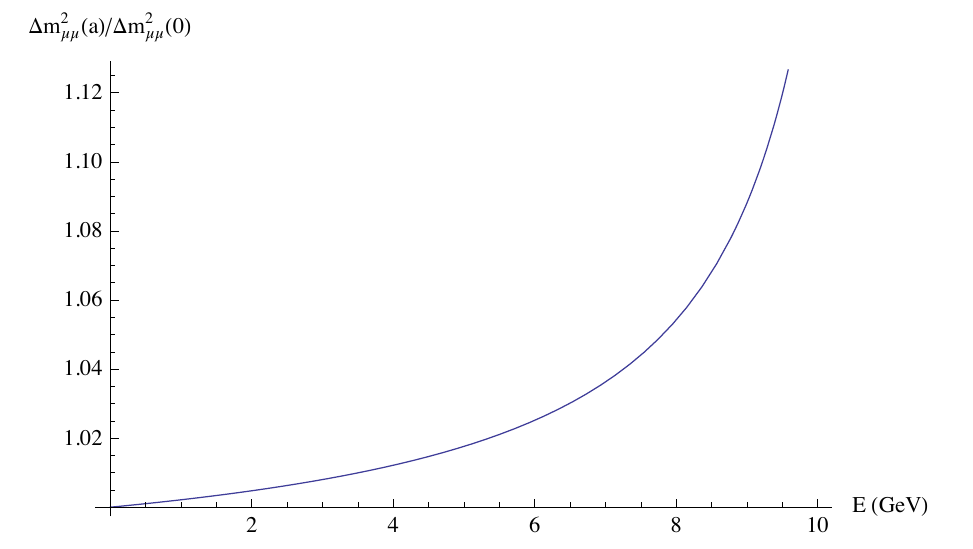}
\end{center}
\caption{
The ratio 
$\Delta m^2_{\mu \mu} (a) / \Delta m^2_{\mu \mu} (0)$ is plotted as a function of $E$ in units of GeV. We take $L=1000$ km. The mixing parameters and the matter density used are the same as in figure~\ref{fig:Dm2-ratio-ee}.
}
\label{fig:Dm2-ratio-mumu}
\end{figure}

Putting aside the problem of $L$-dependence of $\Delta m^2_{\mu \mu} (a)$, we examine its matter potential dependence by examining energy dependence of $\Delta m^2_{\mu \mu} (a) / \Delta m^2_{\mu \mu} (0)$. From figure~\ref{fig:Dm2-ratio-mumu}, one can see that the matter effect correction to $\Delta m^2_{\mu \mu}$ is only a few \% in the ``safe'' region $E \lsim 7$ GeV.

As will be commented at the end of appendix~\ref{sec:hesitation-theorem} the $\nu_{\tau}$ appearance probability $P(\nu_{\tau} \rightarrow \nu_{\tau} )$ can be obtained from $P(\nu_{\mu} \rightarrow \nu_{\mu} )$ by the transformation $c_{23} \rightarrow - s_{23}, s_{23} \rightarrow c_{23}$. Therefore, $\Delta m^2_{\tau \tau} (a)$ can be obtained by the same transformation from $\Delta m^2_{\mu \mu} (a)$.

\subsection{Matter potential dependence of $\Delta m^2_{ee}$ and $\Delta m^2_{\mu \mu}$} 

The matter potential dependence of the effective $\Delta m^2$ is very different between $\Delta m^2_{ee} (a)$ and $\Delta m^2_{\mu \mu} (a)$, as shown in the previous sections. In contrast to the strong matter dependence of $\Delta m^2_{ee} (a)$, $\Delta m^2_{\mu \mu} (a)$ shows only a weak dependence on the matter potential $a$. 

To understand the difference, in particular, the weak matter effect in $P(\nu_{\mu} \rightarrow \nu_{\mu} )$, we derive in appendix~\ref{sec:hesitation-theorem} a general theorem about the matter potential dependence of the various oscillation probabilities, which may be called as the ``matter hesitation theorem''. It states that the matter potential dependent terms in the oscillation probabilities $P(\nu_{\alpha} \rightarrow \nu_{\beta})$ ($\alpha, \beta = e, \alpha, \tau$) receive the suppression factors of at least $s^2_{13}$, or $\epsilon s_{13}$, or $\epsilon^2$, where $\epsilon \equiv \Delta m^2_{21} / \Delta m^2_{31}$ as defined in (\ref{epsilon-def}). That is, the matter effect hesitates to come in before computation reaches to these orders. Given the small values of the parameters, $s^2_{13} \simeq 0.02$, or $\epsilon s_{13}  \simeq 4.5 \times 10^{-3}$, or $\epsilon^2 \simeq 10^{-3}$, the theorem strongly constrains the matter potential dependence of the oscillation probabilities. Our discussion simply generalizes the similar one given in ref.~\cite{Kikuchi:2008vq}. 

Let us apply the matter hesitation theorem to $P(\nu_{\mu} \rightarrow \nu_{\mu} )$, whose expression is given (though in a decomposed way) in (\ref{Palpha-alpha-2flavor3}) with (\ref{matching2}). It reveals the feature of large vacuum term corrected by the suppressed matter effect terms, as dictated by the theorem. Then, we immediately understand the reason why the matter effect dependent terms in $\Delta m^2_{\mu \mu} (a)$, the second line in (\ref{Dm2-eff-mumu-matt}), are suppressed with the factors either $s^2_{13}$ or $\epsilon s_{13}$, explaining its smallness and the weak energy dependence of $\Delta m^2_{\mu \mu} (a)$. 

Then, a question might arises: Given the universal (channel independent) suppression of the matter effect why it can produce a strong modification to $\Delta m^2_{ee}$ in vacuum? Look at first (\ref{Pee-matter}) to notice that all the terms in $1 - P(\nu_{e} \rightarrow \nu_{e} )$ is matter dependent, and they are all equally suppressed by $s^2_{13}$ or by smaller factors. Therefore, the theorem itself is of course valid. But, since all the terms are universally suppressed by small factors, the suppression itself does not tell us how strongly the matter potential affects $1 - P(\nu_{e} \rightarrow \nu_{e} )$. It turned out that the matter effect significantly modifies $1 - P(\nu_{e} \rightarrow \nu_{e} )$, as we have leaned in section~\ref{sec:Dm2-ee-matt}. The feature stems from the structure of matter Hamiltonian $\propto \text{diag} [a, 0, 0]$, which allows $\nu_{e}$ to communicate directly with the matter potential. Even after including the three flavor effect, this feature dominates. 

\section{Effective two-flavor approximation of appearance probability in vacuum }
\label{sec:P-beta-alpha-vac}

In this paper, so far, we have discussed the validity of the concept of effective two-flavor form of the disappearance probability, and the associated effective $\Delta m^2$ in vacuum and in matter. Do these concepts have validities also for the appearance probability? Since we have questioned the validity of the notion of effective $\Delta m^2$ in matter our discussion in this section primarily deal with the possible validity of effective appearance $\Delta m^2$ in vacuum. 

The appearance probability $P(\nu_{\beta} \rightarrow \nu_{\alpha})$ ($\beta \neq \alpha$) in vacuum can be written to order $\epsilon$ in the form 
\begin{eqnarray} 
P(\nu_{\beta} \rightarrow \nu_{\alpha}) =
4 A_{31}^{\beta \alpha} \sin^2 \Delta_{31} +
4 A_{32}^{\beta \alpha} \sin^2 \Delta_{32} +
8 J_{r} c^2_{13} \sin \delta 
\sin \Delta_{21} \sin \Delta_{31} \sin \Delta_{32} 
\label{P-beta-alpha-vac}
\end{eqnarray}
where the sign of CP-odd term in (\ref{P-beta-alpha-vac}) is normalized for $\beta=e$ and $\alpha=\mu$. The coefficients $A_{31}^{\beta \alpha}$ etc are given in table~\ref{tab:coefficient-A}.
%
\begin{table}[h!]
\begin{center}
\caption{
The coefficients $A_{31}^{e \mu}$ etc. used in eq.~(\ref{P-beta-alpha-vac}) are tabulated. The similar expressions for other channel, e.g., $A_{31}^{e \tau}$ can be obtained by the appropriate transformation from $A_{31}^{e \mu}$. See e.g., ref.~\cite{Kikuchi:2008vq}.
}
\label{tab:coefficient-A} 
\begin{tabular}{c|c}
\hline 
\hline 
$A_{31}^{e \mu}$ & 
$c^2_{12} s^2_{23} c^2_{13} s^2_{13} + c^2_{13} J_r \cos \delta$ \\
\hline 
$A_{32}^{e \mu}$ & 
$s^2_{12} s^2_{23} c^2_{13} s^2_{13} - c^2_{13} J_r \cos \delta$ \\
\hline 
$A_{31}^{\mu \tau}$ & 
$c^2_{23} s^2_{23} c^2_{13} \left( s^2_{12} - c^2_{12} s^2_{13} \right) - \cos 2 \theta_{23} c^2_{13} J_r \cos \delta$ \\
\hline 
$A_{32}^{\mu \tau}$ & 
$c^2_{23} s^2_{23} c^2_{13} \left( c^2_{12} - s^2_{12} s^2_{13} \right) + \cos 2 \theta_{23} c^2_{13} J_r \cos \delta$ \\
\hline \hline 
$B_{e \mu}$ & 
$s^2_{23} c^2_{13} s^2_{13} + 2 c^2_{13} J_r \sin \delta \Delta_{21}$ \\
\hline
$B_{\mu \tau}$ & 
$c^2_{23} s^2_{23} c^4_{13} + 2 c^2_{13} J_r \sin \delta \Delta_{21}$ \\
\hline \hline 
\end{tabular}
\end{center}
\end{table}

In complete analogy to the case of survival probability we define the effective $\Delta m^2$ for appearance channel, $\Delta m^2_{\beta \alpha} \equiv \Delta m^2_{\beta \alpha} (0)$, removing ``$(0)$'' (which signals that it is in vacuum) since all the effective $\Delta m^2$ in this section are in vacuum, as 
\begin{eqnarray} 
\Delta m^2_{31} = \Delta m^2_{\beta \alpha} + s_{\beta \alpha} \Delta m^2_{21}. 
\label{Dm2-beta-alpha}
\end{eqnarray}
The effective two-flavor form 
\begin{eqnarray} 
P(\nu_{\beta} \rightarrow \nu_{\alpha}) = 
4 B_{\beta \alpha} \sin^2 \Delta_{\beta \alpha}, 
\label{effective-2flavor-app}
\end{eqnarray}
where $\Delta_{\beta \alpha} \equiv \frac{\Delta m^2_{\beta \alpha} L}{4E}$, is obtained by requiring that the order $\epsilon$ terms that arise from the first two terms in (\ref{P-beta-alpha-vac}) cancel out. Notice that to order $\epsilon$ the CP-odd term in (\ref{P-beta-alpha-vac}) merely renormalizes the coefficient of the effective two-flavor form. The cancellation condition determines $s_{\beta \alpha}$ as 
\begin{eqnarray}
s_{e \mu} 
&=& s^2_{12} -
\frac{ J_r \cos \delta 
}{
s^2_{23} s^2_{13} }, 
\nonumber \\
s_{\mu \tau} 
&=& c^2_{12} + 
\cos 2 \theta_{12} \tan^2 \theta_{13} + 
\frac{ \cos 2\theta_{23} }{ c^2_{23} s^2_{23} c^2_{13} }
J_r \cos \delta. 
\label{s-beta-alpha-vac}
\end{eqnarray}
The resultant coefficients $B_{\beta \alpha}$ for the two-flavor form (\ref{effective-2flavor-app}) are also tabulated in table~\ref{tab:coefficient-A}. 
Notice that $s_{e \mu}$ cannot be expanded in terms of $s_{13}$, because $P(\nu_{e} \rightarrow \nu_{\mu}) =0$ at $s_{13}=0$. The second term of $s_{e \mu}$ signals discrepancy between disappearance $\Delta m^2_{ee}$ and appearance $\Delta m^2_{e \mu}$ in vacuum. Similarly, the difference between $s_{\mu \tau}$ in (\ref{s-beta-alpha-vac}) and $s_{\mu}$ in (\ref{s-mu-vac}) indicate the discrepancy between disappearance and appearance effective $\Delta m^2$. If expanded in terms of $s_{13}$ and keeping to order $\epsilon s_{13}$, $s_{\mu}$ in (\ref{s-mu-vac}) and $s_{\mu \tau}$ in (\ref{s-beta-alpha-vac}) are given by 
$s_{\mu} = c^2_{12} - \frac{ 2 }{ c^2_{23} } J_r \cos \delta$ and $s_{\mu \tau} 
= c^2_{12} + \frac{ \cos 2\theta_{23} }{ c^2_{23} s^2_{23} } J_r \cos \delta$, respectively. They lead to the effective $\Delta m^2$ in disappearance and appearance channels as (without expanding by $s_{13}$ in the $\nu_e$ channel)
\begin{eqnarray}
\Delta m^2_{e e} &=& 
\Delta m^2_{31} - s^2_{12} \Delta m^2_{21}, 
\nonumber \\
\Delta m^2_{e \mu} &=& 
\Delta m^2_{31} - \left( s^2_{12} - \frac{ J_r \cos \delta }{ s^2_{23} s^2_{13} } \right) \Delta m^2_{21}, 
\nonumber \\
\Delta m^2_{\mu \mu} &=& 
\Delta m^2_{31} - \left( c^2_{12} - \frac{ 2 }{ c^2_{23} } J_r \cos \delta \right) \Delta m^2_{21}, 
\nonumber \\
\Delta m^2_{\mu \tau} &=& 
\Delta m^2_{31} -  
\left( c^2_{12} + \frac{ \cos 2\theta_{23} }{ c^2_{23} s^2_{23} } J_r \cos \delta \right) \Delta m^2_{21}.  
\label{Dm2-vac-summary}
\end{eqnarray}

To summarize the results of discussion in this section, we have shown that the effective two-flavor form of appearance probabilities in vacuum can be defined  with suitably defined effective $\Delta m^2$ in parallel to those in disappearance channels. However, the notable feature is that the appearance effective $\Delta m^2$ is different from the corresponding disappearance effective $\Delta m^2$ by an amount of order $\epsilon$ which is proportional to $J_r \cos \delta$. 

What is the meaning of this result? Is it natural to expect that the difference is only the term proportional to $J_r \cos \delta$? The effective $\Delta m^2$ is defined in such a way that it absorbs certain effects which come from the genuine three-flavor properties of the oscillation probability, thereby making it the ``two-flavor'' form. The $\delta$ dependence, not only $\sin \delta$ but also $\cos \delta$, is one of the most familiar examples of such three-flavor effect \cite{Asano:2011nj}. The relative importance of $\cos \delta$ term is different between the probabilities in the appearance and disappearance channels, and it is reflected to the difference the effective $\Delta m^2$. Thus, the feature we see in (\ref{Dm2-vac-summary}) is perfectly natural. The fact that the difference between the appearance and disappearance effective $\Delta m^2$ consists only of $\cos \delta$ term is due to our restriction to first order in $s_{13}$.

One may ask if the similar discussion can go through for the effective two-flavor form of appearance probabilities in matter. The answer to this question is far from obvious to the present author. Even in the simpler case of $\nu_e$ related channels in which $P(\nu_e \rightarrow \nu_e)$ has the two-flavor form (see eq.~(\ref{Pee-matter-SC})) it is unlikely that $P(\nu_e \rightarrow \nu_\mu)$ can be written as the similar two-flavor form under the framework of $\epsilon$ perturbation theory. If one looks at eq.~(3.14) in \cite{Minakata:2015gra}, $P(\nu_e \rightarrow \nu_\mu)$ has a structure similar to (\ref{P-beta-alpha-vac}), but all the eigenvalue differences are of order unity. For more about this point see the discussion in the next section. 

\section{Conclusion and Discussion}
\label{sec:conclusion}

In this paper, we have discussed a question of whether the effective two-flavor approximation of neutrino survival probabilities is viable in matter. We gave an affirmative answer using the perturbative treatment of the oscillation probabilities to order $\epsilon^2$ (to order $\epsilon$ in $\nu_{\mu}$ channel) with the small expansion parameters $\epsilon = \Delta m^2_{21} / \Delta m^2_{31}$ assuming $s_{13} \sim \sqrt{\epsilon}$. It allows us to define the effective $\Delta m^2_{\alpha \alpha} (a)$ ($\alpha = e, \mu, \tau$) in matter in an analogous fashion as in vacuum. However, the resultant expression of $\Delta m^2_{\alpha \alpha} (a)$ poses the problem. 

In neutrino oscillation in vacuum the oscillation probability is a function of $L/E$. However, the effective $\Delta m^2_{\alpha \alpha}$ ($\alpha = e, \mu, \tau$) is defined such that it depends neither on $E$, nor $L$. It is a combination of the fundamental parameters in nature. 
In matter, however, $\Delta m^2_{\alpha \alpha} (a)$ becomes $E$ dependent, which may be permissible because it comes from the Wolfenstein matter potential $a \propto E$. In fact, in $\nu_{e}$ disappearance channel, we have a sensible definition of $\Delta m^2_{ee} (a)$, eq.~(\ref{Dm2ee-matter-SC}), in leading order in the renormalized helio-perturbation theory. However, in the $\nu_{\mu}$ disappearance channel, we have observed that $\Delta m^2_{\mu \mu} (a)$ (and $\Delta m^2_{\tau \tau} (a)$) is $L$ dependent, although we did the same construction of the effective two-flavor form of the survival probability as in the $\nu_{e}$ channel. It casts doubt on whether it is a sensible quantity to define. Certainly, it is an effective quantity which results when we seek the two-flavor description of the three-flavor oscillation probabilities in our way. Nonetheless, the basic three-flavor nature of the phenomena seems to prevent such two-flavor description in the $\nu_{\mu}$ channel. Thus, the effective $\Delta m^2$ in matter does not appear to have any fundamental physical significance. 

One may ask: Is $L$ dependence of $\Delta m^2_{\mu \mu} (a)$ an artifact of the perturbative treatment of $s_{13}$ dependence? We strongly suspect that the answer is {\em No}, though it is very difficult to give an unambiguous proof of this statement at this stage. A circumstantial evidence for the above answer is that we have used the same method of formulating the effective two-lavor approximation in both $\nu_{e}$ and $\nu_{\mu}$ channels. In contrast to $\Delta m^2_{\mu \mu} (a)$, $\Delta m^2_{ee} (a)$ does not have problem of $L$ dependence. It should also be emphasized that the perturbative expression of $\Delta m^2_{ee} (a)$ can be obtained from the ``non-perturbative'' expression (\ref{Dm2ee-matter-SC}) derived by using the renormalized helio-perturbation theory. Notice that the expression (\ref{Dm2ee-matter-SC}) is free from any ``singularity'' at $r_{A}=1$. Therefore, we have no reason to doubt validity of our method used to formulate the effective two-lavor approximation, which treats both the $\nu_{e}$ and $\nu_{\mu}$ channels in an equal footing. 

Putting aside the above conceptual issue, we have examined the matter effect dependences of $\Delta m^2_{ee} (a)$ and $\Delta m^2_{\mu \mu} (a)$. In fact, they are very different. It produces a strong linear energy dependence for $\Delta m^2_{ee} (a)$, whereas $\Delta m^2_{\mu \mu} (a)$ only has a weak energy dependence with magnitude of a few \% level. We expect that the effect of deviation of $\Delta m^2_{ee} (a)$ from the vacuum expression can be observed in a possible future super-LBL experiments, such as neutrino factory, with $\nu_{e}$ detection capability. 

We have also examined the question of whether the similar effective two-flavor form of appearance probability exists with the ``appearance effective $\Delta m^2$''. We have shown that in vacuum it does under the same framework of expanding to order $\epsilon$. We have observed that the effective $\Delta m^2$ in disappearance and appearance channels in vacuum differ by the terms proportional to $\epsilon J_r \cos \delta$. In matter, the effective two-flavor form is very unlikely to exist in the current framework. 

A remaining question would be: What is the meaning of finding, or not finding, the effective two-flavor description of the neutrino oscillation probability in vacuum and in matter? In vacuum we have shown that to order $\epsilon$ such description is tenable in both the appearance and the disappearance channels. It is not too surprising because we restrict ourselves into the particular kinematical region at around the first oscillation maximum, and are expanding by $\epsilon$ to first order, whose vanishing limit implies the two-flavor oscillation. What may be worth remarking is that the effective two-flavor description does not appear to work  in matter under the same approximation as used in vacuum. Nothing magical happens here. Due to the eigenvalue flow as a function of the matter potential all the three eigenvalue differences becomes order unity, and the $\epsilon \rightarrow 0$ limit does not render the system the two-flavor one. 

Finally, in an effort to understand the reasons why the matter potential dependence of $\Delta m^2_{\mu \mu} (a)$ is so weak, we have derived a general theorem which states that the matter potential dependent terms in the oscillation probability are suppressed by a factor of one of $s^2_{13}$, or $\epsilon s_{13}$, or $\epsilon^2$. See appendix~\ref{sec:hesitation-theorem}.

\appendix

\section{Matter Hesitation Theorem}
\label{sec:hesitation-theorem}

\subsection{Statement of the theorem and commentary}
\label{sec:commentary}

In this appendix, we derive the ``matter hesitation theorem'' which states that the matter potential dependent terms in the oscillation probabilities $P(\nu_{\alpha} \rightarrow \nu_{\beta})$ ($\alpha, \beta = e, \alpha, \tau$) receive the suppression factors of at least one of $s^2_{13}$, $\epsilon s_{13}$, or $\epsilon^2$, where $\epsilon \equiv \Delta m^2_{21} / \Delta m^2_{31}$.\footnote{
It should be noticed that exploitation of the perturbative framework for proving the theorem (see below) with expansion parameter $\epsilon$ precludes the possibility of applying it to the solar MSW resonance \cite{Wolfenstein:1977ue,Mikheev:1986gs}. 
}
That is, the matter effect hesitates to come in before computation goes to these orders in $s_{13}$ or in $\epsilon$. It generalizes the discussion given in ref.~\cite{Kikuchi:2008vq} (in particular, its arXiv version 1) which, to our knowledge, first raised the issue of matter hesitation in a systematic way. The discussion in ref.~\cite{Kikuchi:2008vq} uses a specific perturbative framework assuming $s_{13} \sim \epsilon$. 
What we should do here is merely giving a separate treatment for $\epsilon$ and $s_{13}$. 
 
The theorem explains, among other things, the reason why it is so difficult to detect the matter effect in LBL accelerator neutrino experiments even when the $\nu_{e}$-related appearance channel ($\nu_{\mu} \rightarrow \nu_{e}$, or its T-conjugate) is utilized. The matter potential dependent terms in the oscillation probabilities are suppressed by the factors $s^2_{13} \simeq 0.02$, or $\epsilon s_{13}  \simeq 4.5 \times 10^{-3}$, or $\epsilon^2 \simeq 10^{-3}$, which range from reasonably small to quite small. Moreover, if baseline distance is only modest, $\sim 1000$ km or so, $r_{A} \simeq 0.18$ at around the first oscillation maximum, leading to a further suppression of the matter effect.  

Unfortunately, it appears that no general discussion about the matter hesitation phenomenon is available in the literature. Therefore, we present below a perturbative ``proof'' of the matter hesitation theorem. It is not quite a proof but just giving instructions on how to compute a few lowest order terms in the expansion parameters $\epsilon$ and $s_{13}$, which however is sufficient to show the validity of the theorem. Our treatment is valid for arbitrary matter profile.

\subsection{Tilde basis}
\label{sec:tilde-basis}

Neutrino evolution in matter can be described in the flavor basis with the Schr\"odinger equation, $ i \frac{d}{dx} \nu = H \nu $ with 
$H = \frac{ 1 }{ 2E } \left[ U~\text{diag} (0, \Delta m^2_{21}, \Delta m^2_{31} )~U^{\dagger} +\text{diag} (a, 0, 0) \right]$, where 
$U$ denotes the MNS matrix and $a$ the matter potential (\ref{matt-potential}). To formulate the perturbation theory, it is convenient to use the tilde-basis $\tilde{\nu} = U_{23}^{\dagger} \nu $ with Hamiltonian $\tilde{H}$ defined by $H = U_{23} \tilde{H} U_{23}^{\dagger}$. The tilde-basis Hamiltonian is decomposed as $ \tilde{H} = \tilde{H}_{0} + \tilde{H}_{1} $, where 
\begin{eqnarray} 
\tilde{H}_{0} (x) &=& 
\Delta 
\left[
\begin{array}{ccc}
r_{A} (x) & 0 & 0 \\
0 & 0 & 0 \\
0 & 0 & 1
\end{array}
\right] 
\label{H0}
\\
\tilde{H}_{1} (x) &=& 
\Delta 
\left[
\begin{array}{ccc}
s^2_{13} & 0 & c_{13} s_{13} e^{ -i \delta} \\
0 & 0 & 0 \\
c_{13} s_{13} e^{ i \delta} & 0 & - s^2_{13} 
\end{array}
\right] 
+ 
\Delta \epsilon \left[
\begin{array}{ccc}
s^2_{12} c^2_{13} & c_{12} s_{12} c_{13} & - s^2_{12} c_{13} s_{13} e^{ -i \delta} \\
c_{12} s_{12} c_{13} & c^2_{12} & - c_{12} s_{12} s_{13} e^{ - i \delta} \\
- s^2_{12} c_{13} s_{13} e^{ i \delta} & - c_{12} s_{12} s_{13} e^{ i \delta} & s^2_{12} s^2_{13} 
\end{array}
\right] 
\nonumber
\\
\label{H1tilde}
\end{eqnarray} 
where 
$\Delta \equiv \frac{ \Delta m^2_{31} }{2E} $
$\epsilon \equiv \frac{ \Delta m^2_{21} }{ \Delta m^2_{31} } $, 
$r_{A} (x) \equiv \frac{ a(x) }{ \Delta m^2_{31} } $. 

Notice that once the $S$ matrix in the tilde basis, $\tilde{S}$, is obtained the $S$ matrix is obtained as $S(L) = U_{23} \tilde{S} (L)  U_{23}^{\dagger} $, or in an explicit matrix form as 
\begin{eqnarray}
&& \hspace{-5mm} S = 
\nonumber \\
&& \hspace{-6mm} 
\left[
\begin{array}{ccc}
\tilde{ S }_{ee} & c_{23} \tilde{ S }_{e \mu} + s_{23} \tilde{ S }_{e \tau} & - s_{23} \tilde{ S }_{e \mu} + c_{23} \tilde{ S }_{e \tau}  \\
c_{23} \tilde{ S }_{\mu e} + s_{23} \tilde{ S }_{\tau e} & c^2_{23} \tilde{ S }_{\mu \mu} + s^2_{23} \tilde{ S }_{\tau \tau} + c_{23} s_{23} ( \tilde{ S }_{\mu \tau} + \tilde{ S }_{\tau \mu} ) & c^2_{23} \tilde{ S }_{\mu \tau} - s^2_{23} \tilde{ S }_{\tau \mu} + c_{23} s_{23} ( \tilde{ S }_{\tau \tau} - \tilde{ S }_{\mu \mu} ) \\
- s_{23} \tilde{ S }_{\mu e} + c_{23} \tilde{ S }_{\tau e} & c^2_{23} \tilde{ S }_{\tau \mu } - s^2_{23} \tilde{ S }_{ \mu \tau} + c_{23} s_{23} ( \tilde{ S }_{\tau \tau} - \tilde{ S }_{\mu \mu} ) & s^2_{23} \tilde{ S }_{\mu \mu} + c^2_{23} \tilde{ S }_{\tau \tau} - c_{23} s_{23} ( \tilde{ S }_{\mu \tau} + \tilde{ S }_{\tau \mu} ) 
\end{array}
\right]. 
\nonumber \\
\label{SbyS-tilde}
\end{eqnarray}
It should be noticed that $S_{e e} = \tilde{S}_{e e}$, and the relationship between $\tilde{ S }$ and $S$ matrix elements closes inside the $2 \times 2$ $\nu_\mu - \nu_\tau$ sub-block. 

\subsection{Proof of the theorem using the interaction representation}
\label{sec:proof}

To prove the matter hesitation theorem let us introduce the $\hat{\nu}$ basis, $\tilde{\nu} = e^{ - i \int^{x}_{0} dx' \tilde{H}_{0} (x')} \hat{\nu}$. The $\hat{\nu}$' obeys the Schr\"odinger equation $i \frac{d}{dx}  \hat{\nu} =  H_{int}  \hat{\nu} $ 
with $H_{int}$ defined as
\begin{eqnarray} 
H_{int} (x) \equiv e^{i \int^{x}_{0} dx' \tilde{H}_{0} (x')}  
\tilde{H}_{1} e^{- i \int^{x}_{0} dx' \tilde{H}_{0} (x')} 
\label{def-Hint}
\end{eqnarray}
It is nothing but the  ``interaction representation''. For bookkeeping purpose, we decompose the Hamiltonian $H_{int}$ into the terms independent and dependent of the solar-atmospheric ratio $\epsilon$, $H_{int} = H_{int}^{\oplus} + H_{int}^{\odot}$: 
%
%
\begin{eqnarray} 
&& H_{int}^{\oplus} (x)  = 
\Delta  \left[
\begin{array}{ccc}
s^2_{13} & 0 &  
c_{13} s_{13} e^{ -i \delta} e^{ - i \Delta \int^{x}_{0} dx' [1 - r_{A} (x') ] }  \\
0 & 0 & 0  \\
c_{13} s_{13} e^{ i \delta} e^{ i \Delta \int^{x}_{0} dx' [1 - r_{A} (x') ] } & 
0 & - s^2_{13} 
\end{array}
\right], 
\label{Hint-atm}
\end{eqnarray}

\begin{eqnarray} 
&& H_{int}^{\odot} (x) = \epsilon \Delta 
\nonumber \\
&\times& 
\left[
\begin{array}{ccc}
s^2_{12} c^2_{13} & 
c_{12} s_{12} c_{13} e^{ i \Delta \int^{x}_{0} dx' r_{A} (x')  } & 
- s^2_{12} c_{13} s_{13} e^{ -i \delta} e^{ - i \Delta \int^{x}_{0} dx' [1 - r_{A} (x') ] }  \\
c_{12} s_{12} c_{13} e^{ - i \Delta \int^{x}_{0} dx' r_{A} (x')  }  & 
c^2_{12} & 
- c_{12} s_{12} s_{13} e^{ -i \delta} e^{ - i \Delta x }   \\
- s^2_{12} c_{13} s_{13} e^{ i \delta} e^{ i \Delta \int^{x}_{0} dx' [1 - r_{A} (x') ] } & 
- c_{12} s_{12} s_{13} e^{ i \delta} e^{ i \Delta x }  & 
s^2_{12} s^2_{13}  
\end{array}
\right]. 
\nonumber \\
\label{Hint-solar}
\end{eqnarray}
The interaction representation Hamiltonian $H_{int}$ has a peculiar feature that there is no matter potential dependence in the $\nu_{\mu} - \nu_{\tau}$ sector as well as in the $\nu_{e}$-$\nu_{e}$ element. 
It is nothing but this feature of $H_{int}$ in (\ref{Hint-atm}) and (\ref{Hint-solar}) that the matter effect is absent, to first order in $s_{13}$ and $\epsilon$, 
in the oscillation probabilities in the $\nu_{\mu} - \nu_{\tau}$ as well as in $P(\nu_{e} \rightarrow \nu_{e} )$. 

To confirm this understanding and find out what happens in the $\nu_{e} \rightarrow \nu_{\mu}$ and $\nu_{e} \rightarrow \nu_{\tau}$ appearance channels, we calculate the $S$ matrix in the tilde basis 
\begin{eqnarray} 
&&\tilde{S} (L) =  
e^{ - i \int^{L}_{0} dx' \tilde{H}_{0} (x')} 
\nonumber \\
&\times&
\left[ 1 + 
(-i) \int^{L}_{0} dx'  H_{int} (x') + 
(-i)^2 \int^{L}_{0} dx'  H_{int} (x') \int^{x'}_{0} dx''  H_{int} (x'')  + \cdot \cdot \cdot
\right]
\label{tilde-S}
\end{eqnarray}
where the ``space-ordered'' form in (\ref{tilde-S}) is essential because of the highly nontrivial spatial dependence in $H_{int}$. 
The elements of $ \tilde{S} (L) $ are given to order $s^2_{13}$ and $\epsilon s_{13}$ as 
\begin{eqnarray} 
\tilde{S} (L)_{ee}  &=& 
 \left[ 1 - i \left( s^2_{13} + \epsilon s^2_{12} c^2_{13} \right) \Delta L  \right] e^{ - i \Delta \int^{L}_{0} dx r_{A} (x) } 
\nonumber \\ 
&-& c^2_{13} s^2_{13} \Delta^2 e^{ - i \Delta \int^{L}_{0} dx r_{A} (x) } 
\int^{L}_{0} dx' e^{ - i \Delta \int^{x'}_{0} dy [ 1 - r_{A} (y) ] } 
\int^{x'}_{0} dz e^{ i \Delta \int^{z}_{0} dy [ 1 - r_{A} (y) ]  }, 
\nonumber \\
\tilde{S} (L)_{e \mu}  &=& 
- i \epsilon c_{12} s_{12} c_{13} \Delta 
e^{ - i \Delta \int^{L}_{0} dx r_{A} (x) } 
\int^{L}_{0} dx e^{ i \Delta \int^{x}_{0} dx' r_{A} (x')  }, 
\nonumber \\
\tilde{S} (L)_{\mu e}  &=& 
- i \epsilon c_{12} s_{12} c_{13} \Delta 
\int^{L}_{0} dx e^{ - i \Delta \int^{x}_{0} dx' r_{A} (x')  }, 
\nonumber \\
\tilde{S} (L)_{e \tau}  &=& 
 - i c_{13} s_{13} e^{ - i \delta} \left( 1 -  \epsilon s^2_{12} \right)  \Delta 
 e^{ - i \Delta \int^{L}_{0} dx r_{A} (x) } 
\int^{L}_{0} dx e^{ - i \Delta \int^{x}_{0} dx' [1 - r_{A} (x') ] } 
\nonumber \\ 
&-& \epsilon s^2_{12} c^3_{13} s_{13} e^{ - i \delta}  \Delta^2  
e^{ - i \Delta \int^{L}_{0} dx r_{A} (x) } 
\int^{L}_{0} dx' 
\int^{x'}_{0} dz e^{ - i \Delta \int^{z}_{0} dy [ 1 - r_{A} (y) ]  }, 
\nonumber \\
\tilde{S} (L)_{\tau e}  &=& 
- i c_{13} s_{13} e^{ i \delta} \left( 1 -  \epsilon s^2_{12} \right)  \Delta 
 e^{ - i \Delta L } 
\int^{L}_{0} dx e^{ i \Delta \int^{x}_{0} dx' [1 - r_{A} (x') ] } 
\nonumber \\ 
&-& \epsilon s^2_{12} c^3_{13} s_{13} e^{ i \delta}  \Delta 
 e^{ - i \Delta L } 
\int^{L}_{0} dx' ( \Delta x' )
e^{ i \Delta \int^{x'}_{0} dy [1 - r_{A} (y) ] }. 
\label{Stilde-element-e}
\end{eqnarray}
\begin{eqnarray} 
\tilde{S} (L)_{\mu \mu}  &=& 
1 - i \epsilon c^2_{12} \Delta L, 
\nonumber \\
\tilde{S} (L)_{\mu \tau}  &=& 
+ i \epsilon c_{12} s_{12} s_{13} e^{ - i \delta}  \Delta 
\int^{L}_{0} dx e^{ - i \Delta x  } 
\nonumber \\
&-& 
\epsilon c_{12} s_{12} c^2_{13} s_{13} e^{ - i \delta} \Delta^2  
\int^{L}_{0} dx' 
 e^{ - i \Delta \int^{x'}_{0} dy r_{A} (y) } 
\int^{x'}_{0} dz e^{ - i \Delta \int^{z}_{0} dy [ 1 - r_{A} (y) ] }, 
\nonumber \\
\tilde{S} (L)_{\tau \mu} &=& 
+ i \epsilon c_{12} s_{12} s_{13} e^{ i \delta}  \Delta e^{ - i \Delta L  } 
\int^{L}_{0} dx e^{ i \Delta x } 
\nonumber \\ 
&-& 
\epsilon c_{12} s_{12} c^2_{13} s_{13} e^{ i \delta} \Delta^2  
\int^{L}_{0} dx' 
 e^{ i \Delta \int^{x'}_{0} dx [ 1 - r_{A} (x) ] } 
\int^{x'}_{0} dz e^{ i \Delta \int^{z}_{0} dy r_{A} (y) }, 
\nonumber \\
\tilde{S} (L)_{\tau \tau}  &=& 
\left[ 1 + i \left( s^2_{13} - \epsilon s^2_{12} s^2_{13} \right) \Delta L \right]
e^{ - i \Delta L  } 
\nonumber \\ 
&-& c^2_{13} s^2_{13} \Delta^2 e^{ - i \Delta L  } 
\int^{L}_{0} dx' e^{ i \Delta \int^{x'}_{0} dy [ 1 - r_{A} (y) ] } 
\int^{x'}_{0} dz e^{ - i \Delta \int^{z}_{0} dy [ 1 - r_{A} (y) ]  }. 
\label{Stilde-element-mutau}
\end{eqnarray}
As a rotation by $U_{23}$ does not mix the $\nu_\mu - \nu_\tau$ sector to the $\nu_e - \nu_\alpha$ ($\alpha = e, \mu, \tau$) sector, knowing the structure of the $\tilde{S} (L)$ matrix is sufficient to prove the matter hesitation theorem. 

We first discuss the $\nu_e$-related sector. The survival probability $P(\nu_{e} \rightarrow \nu_{e} )$ can be obtained as $\vert \tilde{S}_{ee} \vert^2$. Square of the first term in $\tilde{S}_{ee}$ contains, to leading order, the terms of order $s^4_{13}$, $\epsilon s^2_{13}$, and $\epsilon^2$, and they are all matter independent terms. Therefore, the lowest-order contribution of $1 - P(\nu_{e} \rightarrow \nu_{e} )$ comes from the interference between the first and the second lines of $\tilde{S}_{ee}$ in (\ref{Stilde-element-e}), and the term is matter potential dependent, and is of order $s^2_{13}$. Hence, the theorem holds, but in a trivial way.\footnote{
This statement requires clarification. Notice that $S_{ee} = \tilde{S}_{ee}$ does have matter effect even in zeroth order in $\epsilon$ because $\tilde{H}_{0}$ has zeroth-order matter effect in $\nu_{e}$ row. It disappears in the survival probability $\vert S_{ee} \vert^2$ only because the matter dependence comes in via the phase factor, as can be seen in the first line in the matrix elements in (\ref{Stilde-element-e}). Therefore, we stress that the absence of the matter effect in the oscillation probability to first order in $\epsilon$ in the $\nu_{e} \rightarrow \nu_{e}$ channel is highly nontrivial. Alternatively, one can argue that exponentiation of the matter potential term (times $i$) must occur based on unitarity \cite{Kikuchi:2008vq}.
}

The appearance probability $P(\nu_{e} \rightarrow \nu_{\mu} )$ (or $P(\nu_{e} \rightarrow \nu_{\tau} )$) can be computed as absolute square of the amplitude which is the superposition of $\tilde{S} (L)_{\mu e}$ and $\tilde{S} (L)_{\tau e}$, as shown in (\ref{SbyS-tilde}). There is no unity term in them, and the amplitudes have terms of order $\sim s_{13}$, $\sim \epsilon$, and $\sim \epsilon s_{13} $, which are all matter potential dependent. By adding them and squaring it one can see that the leading order terms in the $\nu_{\mu}$ (or $\nu_{\tau}$) appearance probability, all matter dependent, are of the order $s^2_{13}$, or $\epsilon s_{13}$, or $\epsilon^2$. That is, there is no matter-dependent order $\epsilon$ terms, which agrees with the statement of the theorem. 

Now, we turn to the $\nu_\mu - \nu_\tau$ sector. The appearance and disappearance channel probabilities in the $\nu_\mu - \nu_\tau$ sector (which we call here $P_{\mu \tau-\text{sect.} }$ collectively) are given by absolute square of the amplitude which is the superposition of $\tilde{S} (L)_{\mu \mu}$, $\tilde{S} (L)_{\mu \tau}$, $\tilde{S} (L)_{\tau \mu}$, and $\tilde{S} (L)_{\tau \tau}$, see (\ref{SbyS-tilde}). Then, one may think that the order $\epsilon$ term in $\tilde{S} (L)_{\mu \mu}$ might produce an order $\epsilon$ terms in $P_{\mu \tau-\text{sect.} }$. But, they are matter-independent vacuum terms, similar to the case of $P(\nu_{e} \rightarrow \nu_{e} )$, and hence no relevance to the statement of the theorem. The matter dependent terms in the amplitudes in the $\nu_\mu - \nu_\tau$ sector are of order either $\epsilon s_{13}$, or $s^2_{13}$, which would produce the terms of these orders by interfering with the order $\epsilon^{0}$ terms in $\tilde{S} (L)_{\mu \mu}$ and $\tilde{S} (L)_{\tau \tau}$. 

This completes the derivation of the matter hesitation theorem, the property that matter effects comes in into the oscillation probabilities only at order $s^2_{13}$, or $\epsilon s_{13}$, or $\epsilon^2$. Of course, the theorem holds in the explicit expressions of the survival probabilities in (\ref{Pee-matter}) and (\ref{Palpha-alpha-2flavor3}) with (\ref{matching2}) (though the latter is not written in a closed form).

Here, we make a supplementally comment that eq.~(\ref{SbyS-tilde}) implies the relationship between the $S$ matrix elements $S_{\tau \tau} = S_{\mu \mu} ( c_{23} \rightarrow - s_{23}, s_{23} \rightarrow c_{23} )$. Therefore, $P(\nu_{\tau} \rightarrow \nu_{\tau} )$ can be obtained from $P(\nu_{\mu} \rightarrow \nu_{\mu} )$ by the same transformation.


Finally, we give a clarifying remark on another aspect of the matter hesitation theorem: 
The theorem implies that there is no matter-dependent order $s_{13}$ terms in the oscillation probabilities. In fact, one can prove a more general statement that there is no such terms with single power of $s_{13}$, matter potential dependent or not, in the oscillation probabilities. 

To understand the point, let us recapitulate the point of the argument given in \cite{Asano:2011nj} for a different purpose, which we try to generalize here in the present context by including $\nu_{\mu} - \nu_{\tau}$ sector. 
We first note that $s_{13}$ and $\delta$ enter into the Hamiltonian through the single variable $z \equiv s_{13} e^{i \delta}$. Therefore, $s_{13}$- and $\delta$-dependences of the oscillation probability $P$ can be written as a power series expansion as 
$P = \sum_{n, m}^{\infty} f_{n m} z^{n} (z^{*})^m$, 
where $f_{n m} = f^{*}_{m n} $ for the reality of $P$. 
Then, $\cos \delta$ and $\sin \delta$ terms in $P$, which comes from the terms $m =n \pm 1$, must have the form $P = K( s^2_{13} ) s_{13} \cos \delta + M( s^2_{13} ) s_{13} \sin \delta$, where $K$ and $M$ are some functions. It means that odd terms in $s_{13}$ must be accompanied with $\cos \delta$ or $\sin \delta$.\footnote{
It is known that in $\nu_{e}$-related sector ($\nu_{\mu} - \nu_{\tau}$ sector) the $\delta$ dependence of the oscillation probabilities is limited to the terms proportional to either $\sin \delta$ or $\cos \delta$ in any ($\sin \delta$, $\cos \delta$, or $\cos 2 \delta$ in symmetric) matter profile \cite{Yokomakura:2002av}. The same discussion as above shows that $\cos 2 \delta$ terms are suppressed at least by $s^2_{13}$. In fact, the term is further suppressed by $\epsilon^2$ \cite{Kimura:2002wd}. 
}
But, then it is shown that these $\delta$-dependent terms (not only sine but also cosine) receive another suppression factor $\epsilon$, the Theorem B in \cite{Asano:2011nj}, indicating their genuine three-flavor nature. Thus, the order $s_{13}$ terms {\em do not} exist in $P$, and lowest order contribution of this type is of order $\epsilon s_{13}$.

\begin{acknowledgments}

The author expresses special thanks to Stephen Parke for suggesting this problem in 2013, since then showing continuous interests in this work with numerous discussions, and, in particular for finally disagreeing with the conclusion in section~\ref{sec:conclusion}, which prompted H.M. to publish this paper. He thanks Hiroshi Nunokawa for discussions which led to introduction of the simple model of energy dependence of $\Delta m^2_{ee} (a)$ in section~\ref{sec:E-dep-Peemin} and for kindly drawing figure~\ref{fig:Pee}. 

He is grateful to Conselho Nacional de Ci\^encia e Tecnologia (CNPq) for support for his visit to Departamento de F\'{\i}sica, Pontif{\'\i}cia Universidade Cat{\'o}lica do Rio de Janeiro in 2013, and to Universidade de S\~ao Paulo for the opportunity of stay under ``Programa de Bolsas para Professors Visitantes Internacionais na USP'', and as a fellow supported by Funda\c{c}\~ao de Amparo \`a Pesquisa do Estado de S\~ao Paulo (FAPESP) under grant 2015/05208-4, which enabled him to spend great time in Brazil until March 2016. 
Finally, special thanks should go to Theory Group of Fermilab for warm hospitalities extended to him during the numerous visits in the course of this work. 


\end{acknowledgments}

\end{document}